\documentclass[journal,usenames,dvipsnames]{IEEEtran}

\IEEEoverridecommandlockouts

\usepackage{graphicx}
\usepackage{ifpdf}
\ifpdf
   \DeclareGraphicsExtensions{.pdf,.png,.jpg,.mps}
   \usepackage{hyperref} 
\else
\fi

\usepackage{mystyle}

\begin{document}

\title{Spotlight: Scalable Transport Layer Load Balancing for Data Center Networks }
\author{\IEEEauthorblockN{
    Ashkan Aghdai\IEEEauthorrefmark{1},
    Cing-Yu Chu\IEEEauthorrefmark{1},
    Yang Xu\IEEEauthorrefmark{1},
    David H. Dai\IEEEauthorrefmark{2},
    Jun Xu\IEEEauthorrefmark{2},
    H. Jonathan Chao\IEEEauthorrefmark{1}}
\IEEEauthorblockA{
    \\ \IEEEauthorrefmark{1}Tandon School of Engineering, New York University, Brooklyn, NY, USA
    %\\ \{ashkan.aghdai,\ cyc391,\ yang,\ chao\}@nyu.edu
    }
\and
\IEEEauthorblockA{
    \\ \IEEEauthorrefmark{2}Huawei Technologies, Santa Clara, CA, USA
    %\\ \{david.h.dai, jun.xu\}@huawei.com
    }
}

\newcommand\copyrighttext{%
  \footnotesize This article is submitted to IEEE Transactions on Cloud Computing and is under review.}
\newcommand\copyrightnotice{%
\begin{tikzpicture}[remember picture,overlay]
\node[anchor=south,yshift=10pt] at (current page.south) {\fbox{\parbox{\dimexpr\textwidth-\fboxsep-\fboxrule\relax}{\copyrighttext}}};
\end{tikzpicture}%
}

\maketitle
\copyrightnotice

\begin{abstract}
	Load Balancing plays a vital role in cloud data centers to distribute traffic among instances of network functions or services.
    State-of-the-art load balancers dispatch traffic obliviously without considering the real-time utilization of service instances and therefore can lead to uneven load distribution and sub-optimal performance.

    In this paper, we design and implement Spotlight, a scalable and distributed load balancing architecture that maintains connection-to-instance mapping consistency at the edge of data center networks.
    Spotlight uses a new stateful flow dispatcher which periodically polls instances' load and dispatches incoming connections to instances in proportion to their available capacity.
    Our design utilizes a distributed control plane and in-band flow dispatching;
    thus, it scales horizontally in data center networks.
    Through extensive flow-level simulation and packet-level experiments on a testbed with HTTP traffic on unmodified Linux kernel, we demonstrate that compared to existing methods Spotlight distributes traffic more efficiently and has near-optimum performance in terms of overall service utilization.
    Compared to existing solutions, Spotlight improves aggregated throughput and average flow completion time by at least 20\% with infrequent control plane updates.
    Moreover, we show that Spotlight scales horizontally as it updates the switches at O(100ms) and is resilient to lack of control plane convergence.
\end{abstract}

\begin{IEEEkeywords}
    software defined networks, scalability, transport layer load balancing, network function virtualization.
\end{IEEEkeywords}

\section{Introduction} 
In a modern cloud data center, a large number of services and network
functions coexist.
On average, 44\% of data center traffic passes through at least one
service~\cite{patel2013ananta}.
Network services scale out with a large number of service instances
to keep up with the ever-growing demand from users.
Data center networks perform load balancing in more than one way.
L3 load balancers select one of the many equal-cost links to route packets to their destination,
while L4 load balancers choose the
serving instances for incoming connections to services.

Services and network functions perform stateful operations on connections.
Consider Intrusion Detection Systems (IDS) as an example.
For an IDS to accurately detect intrusive connections, it should process
the content of a connection as a whole and not on a per-packet basis since
malicious content may be spread across multiple packets.
In other words, judging by individual packets, an IDS cannot reliably decide
whether or not the content is malicious.
Therefore, once a load balancer chooses an IDS instance for a particular
connection, \emph{all} packets of that connection should be processed by the
same IDS instance.
This requirement is referred to as Per-Connection Consistency
(PCC)~\cite{miao2017silkroad}. 
Violating PCC may result in malfunction, connection interruption, or increased
end-to-end latency that degrade the quality of service considerably.

PCC requirement reduces the load balancing problem to the distribution of new
connections among service instances.
A PCC load balancer can be viewed from two aspects:
\begin{enumerate}
    \item \textbf{Maintenance of PCC:} How does the load balancer assure that flows
        are consistently directed to their serving instances?
        This question signifies the architecture and implementation of the
        load balancer.
        Therefore, the answer to this question affects the scalability and
        practicality of the load balancer.
    \item \textbf{Flow Dispatching:} Which instance serves an incoming
        connection?
        The answer to this question determines how well the load balancer
        utilizes its instances.
        Inefficient flow dispatching leads to performance degradation as a result of
        overwhelming some service instances while others are under-utilized.
\end{enumerate}

Data centers relied on dedicated load balancers~\cite{
    netscaler,f5,nginx} to maintain PCC.
Dedicated load balancers route flows through a middlebox that chooses
the serving instances.
While routing all of the traffic through a middlebox simplifies the maintenance
of PCC, it quickly becomes a performance bottleneck as cloud services
scale out.
Distributed load balancers eliminate the performance bottleneck and enable load balancing
to scale out at the same pace as cloud services.
Modern data centers use distributed L4 load balancing schemes.
Some solutions~\cite{patel2013ananta,gandhi2015duet,miao2017silkroad}
use Equal Cost Multipath Routing (ECMP), while others~\cite{
    eisenbud2016maglev,araujo2018balancing,olteanu2018stateless}
use various forms of consistent hashing~\cite{karger1997consistent}
to dispatch flows.

Stateless flow dispatchers such as ECMP and consistent hashing do not
take the real-time utilization of service instances into account and distribute
an equal number of connections among them.
Connections' size distribution is heavy-tailed~\cite{greenberg2009vl2} and
instances may not have a uniform processing power.
Therefore, stateless flow dispatching may lead to uneven utilization of
service instances, which is highly reminiscent of the link utilization discrepancies that were
observed in stateless L3 load balancers~\cite{al2010hedera}.
That problem was the culprit to substantial bandwidth losses at data centers
and led to the development of stateful L3 load balancers~\cite{
    alizadeh2014conga,katta2016hula,katta2017clove} that maximize the aggregated
bandwidth of data center networks by prioritizing least congested links.
Although it is possible to assign static weights to DIPs and implement Weighted Cost Multipath routing (WCMP)~\cite{zhou2014wcmp}, stateless solutions cannot update the weights on the go.

Inspired by the evolution of L3 load balancers, we question the efficiency
of stateless flow dispatching for L4 load balancing.
In this paper we show that stateless flow dispatchers are indeed the cause of
significant throughput losses in services with many serving instances.
Motivated by this observation, we design a stateful flow dispatcher to
distribute traffic among serving instances efficiently and maximize the aggregated 
service throughput.
Using the proposed flow dispatcher, we implement a distributed load balancer
that satisfies the PCC requirement and scales horizontally in data center networks.

\subsection{Contributions}
\subsubsection {Design and implementation of Adaptive Weighted Flow Dispatching (AWFD)}
AWFD is our proposed flow dispatching algorithm that distributes connections
among instances in proportion to instances' available capacity.
Unlike ECMP and consistent hashing, AWFD is stateful; it periodically polls
instances' available capacity to classify them into different priority classes.
Load balancers use priority classes to assign new connections to instances.
Our simulations using backbone ISP traffic traces as well as synthesized heavy-tail distribution, show that for a service with 100 instances,
AWFD with $O(100ms)$ polling interval and 4 priority
classes yields a near-optimum aggregated service throughput.

\subsubsection{Design and implementation of Spotlight}
Spotlight is a platform that enables the scalable and PCC-compliant
implementation of AWFD at the edge of data center networks.
Spotlight estimates instances available capacity and uses this
information to run the AWFD algorithm.
As a Software Defined Networking (SDN) application, Spotlight implements a
distributed control plane to push AWFD priority classes to edge switches.
Edge switches use priority classes to dispatch incoming connections
to service instances in the data plane.
In-band flow dispatching eliminates the pressure on the control plane and allows
Spotlight to scale horizontally.
Moreover, Spotlight is transparent to applications and does not require any modification
at service instances' applications or operating systems.
We have implemented Spotlight on a small scale testbed; in our testbed, Spotlight
load balances HTTP requests to 16 instances that run unmodified Apache~\cite{apache} web server on top of unmodified Linux kernel.
HTTP requests's size distribution is derived from traffic traces from a production data center network.
Our testbed results show that using $O(100ms)$ polling interval, our solution
improves the aggregated throughput and average flow completion time by at least 20\%
compared to stateless ECMP/WCMP-based solutions.

\subsubsection {Providing thorough insights into the scalability of Spotlight}
We explore how Spotlight handles potential inconsistencies in the control plane
and show that it is highly resilient to loss of control plane messages.
We also show that Spotlight generates an insignicant amount of control plane traffic
for load balancing a multi Terabit per second service.

The rest of the paper is organized as follows.
\S\ref{motivationSec} reviews the load balancing problem in fine detail.
\S\ref{mALCSec} explores existing flow dispatchers, presents their weaknesses, 
and proposes AWFD.
\S\ref{SpotlightSec} presents Spotlight.
\S\ref{evalSec} evaluates AWFD and Spotlight using flow-level simulations and
packet-level experiments on a testbed, respectively.
\S\ref{relatedSec} reviews related works in this area.
Finally, \S\ref{conclusionSec} concludes the paper.

\section{Motivation and Background}\label{motivationSec}
\subsection{Terminology}

\begin{table}[t]
    \caption{Notations} \label{tab:Notation} 
    \scriptsize
    \centering
    \begin{tabular}{ | c | c |  }
        \hline
        \textbf{Term} & \textbf{Definition}\\ \hline
        $j$ & VIP index\\ \hline
        $i$ & DIP index\\ \hline
        $N^j$ & Number of instances for $j$th VIP\\ \hline
        $f_i^j$ & $i$th instance of $j$th VIP\\ \hline
        $C_i^j$ & Capacity of $f_i^j$\\ \hline
        $U_i^j$ & Utilization of $f_i^j$\\ \hline
        $L_i^j = U_i^jC_i^j$ & Load at $f_i^j$\\ \hline
        $A_i^j = (1-U_i^j)C_i^j$ & Available capacity of $f_i^j$\\ \hline
        $p[f_i^j]$ & Probability of assigning a new flow to $f_i^j$ \\ \hline
        $C^j = \sum_{i=1}^{N^j}{C_i^j}$ & Capacity of VIP $j$\\ \hline
        $T^j = \sum_{i=1}^{N^j}{U_i^jC_i^j}$ & Aggregated throughput of VIP $j$\\ \hline
        $\Omega^j = T^j/C^j$& Service Utilization for VIP $j$\\
        \hline
    \end{tabular}
\end{table}

In data centers, services or network 
functions are usually assigned Virtual IP addresses (\emph{VIP}).
Each VIP has many instances that collectively perform the network function;
instances are uniquely identified by their Direct IP addresses (\emph{DIP}).
An L4 Load Balancer (LB) distributes the VIP traffic among DIPs by assigning
connections to DIPs; this assignment process is also referred to as
Flow Dispatching or connection-to-DIP mapping.
The connection-to-DIP mapping should be consistent: i.e., all packets of a
connection\footnote{In this paper the terms connection and flow are used
interchangeably.} should be processed by the same DIP (\emph{meeting PCC
requirement}).
We refer to the mapping between active connections and their DIPs as the state
of the load balancer.

Table~\ref{tab:Notation} summarizes the notation that is used throughout the
paper.

\subsubsection{Data Center Networks' Edge}
We refer to networking equipment that meets either of the following conditions
as an edge device:
\begin{enumerate}[label=(\Alph*)]
    \item A device that connects a physical host or a guest VM to the network,
        such as a top of the rack (ToR) switch, hypervisor virtual switch, or
        network interface card.
    \item A device that connects two or more IP domains.
        For example, data centers' border gateway connects them to an IP exchange
        (IPX) or an autonomous system (AS).
\end{enumerate}

Under this definition, a packet may traverse through \emph{many} edges in its
lifetime.
State-of-the-art programmable network switches~\cite{bosshart2013forwarding,tofino}
can be implemented as the first category of edge definition.
It is also possible to deploy programmable switches in place of or as the next hop of border
gateways to enable programmabality at the gateway level.

Modern networks move applications towards the edge and leave
high-speed packet forwarding as the primary function of the core networks.
Examples of this trend include Clove~\cite{katta2017clove} and
Maglev~\cite{eisenbud2016maglev} in data center networks,
mobile edge cloud (MEC) implementations~\cite{hu2015mobile} in mobile networks,
and the deployment of NFV infrastructure (NFVI) at ISPs~\cite{att2013domain}.

\subsection{Load Balancing Architecture}
\begin{figure}[t]
    \centering
    \includegraphics[width=\linewidth]{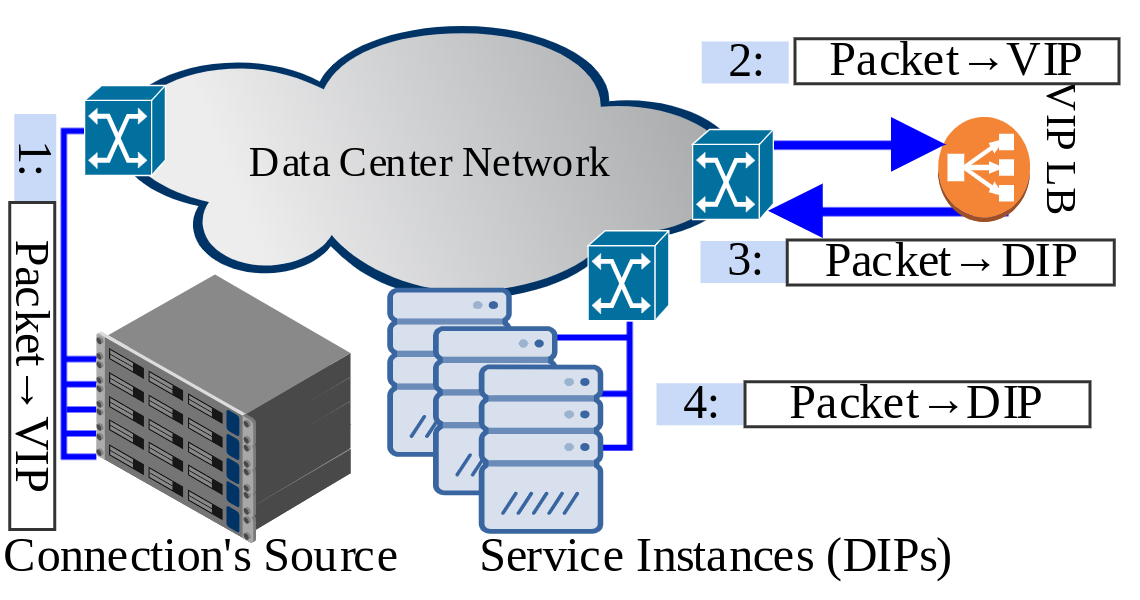}
    \caption{Dedicated load balancing.}\label{dediLB}
\end{figure}

Traditionally, data center operators used a dedicated load balancer for each
VIP.
In this architecture, as illustrated in Figure~\ref{dediLB}, a hardware
device is configured as VIP load balancer.
\emph{All} traffic to the VIP pass through the dedicated load balancer which
uses ECMP to dispatch flows.
Therefore, the dedicated load balancer is a single point of failure as well as
a potential performance bottleneck for its respective service.
Dedicated load balancing is a scale-up architecture utilizing expensive high-capacity hardware.
In this architecture, the load balancer is typically deployed at the core of
the network to handle incoming traffic from the Internet as well as the
intra-data center traffic.
As a result, it adds extra hops to connections' data path.
The advantage of this dedicated architecture is its simplicity since the load
balancing state is kept on a single device and could be easily backed up or
replicated to ensure PCC in case of failures.

\begin{figure}[t]
    \centering
    \includegraphics[width=\linewidth]{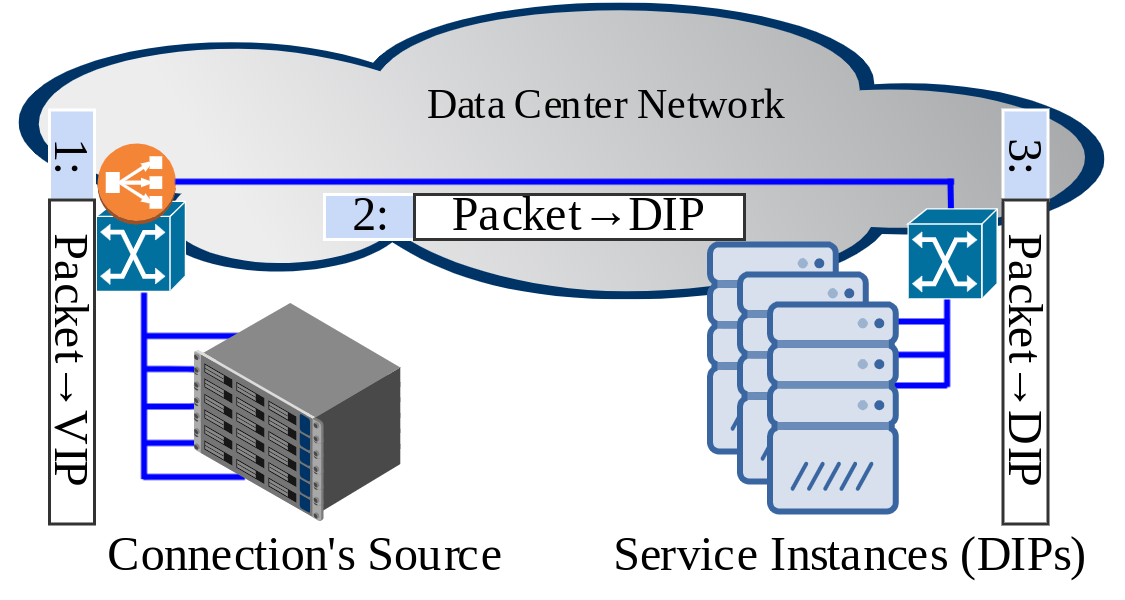}
    \caption{Distributed load balancing at source.}\label{distLB}
\end{figure}

Modern data centers, on the other hand, rely on distributed load
balancing~\cite{eisenbud2016maglev}.
In this architecture, many hardware or software load balancers handle the
traffic for one or many VIPs.
Distributed load balancing is a scale-out architecture that relies on commodity
devices to perform load balancing.
Compared to the dedicated architecture, distributed load balancing can deliver
a higher throughput at a lower cost.
The distributed load balancing on the source side is depicted in
Figure~\ref{distLB}.
Distruted load balancers resolve VIP to DIP for incoming packets.
Compared to dedicated solutions, connections' data path has fewer hops.
This architecture offloads the load balancing function to edge devices.

The most challenging issue in designing distributed load balancers is the
partitioned state that is distributed across multiple devices.
Ensuring PCC poses a challenge in this architecture since load balancers
are prone to failure.
To solve this issue, \cite{eisenbud2016maglev} proposes to use consistent
hashing~\cite{karger1997consistent}.
Consistent hashing allows the system to recover the lost state from failing
devices and thus guarantees PCC.

\subsection{Problem Statement and Motivation}
We focus on dynamic load balancing where connection information such as size,
duration, and rate are \emph{not} available to the load balancer.
An efficient L4 load balancer distributes incoming connections among DIPs
such that:
\begin{enumerate}
    \item The connection-to-DIP mapping remains consistent (PCC).
    \item For each VIP, the aggregated throughput is maximized.
\end{enumerate}

Existing L4 load balancers put much emphasis on meeting the PCC requirement
and treat efficient load distribution as a secondary objective.
A substantial drawback to all of the existing solutions is that they aim to distribute
an equal number of connections among DIPs using ECMP or consistent hashing.
In \S~\ref{mALCSec}, we show that this objective does not maximize the aggregated
throughput of services ($T^j$).

Our primary motivation in designing Spotlight is to maximize $T^j$,
on top of meeting the PCC requirement.

\section{Flow Dispatching}\label{mALCSec}

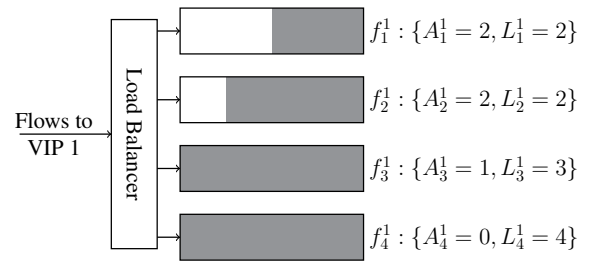
\begin{figure}[b]
    \centering
    \pagenumbering{gobble}

\begin{tikzpicture}
    
    %\fill[Cyan] (0, 0) rectangle(1.5, 3);
    %\draw[thick] (0, 0) rectangle(1.5, 3) node[pos=0.5] {
    %    \begin{tabular}{c}
    %        $f_1^1$ \\
    %        \ \\
    %        $A_1^1=0$ \\
    %        $L_1^1=4$
    %    \end{tabular}
    %};
    \fill[Gray] (2, 0) rectangle(6, 1);
    \draw[thick] (2, 0) rectangle(6, 1);
    \node[right] at (6, 0.5) {\Large
        $f_4^1 : \{A_4^1=0, L_4^1=4\}$
    };

    \fill[Gray] (2, 1.5) rectangle(6, 2.5);
    \draw[thick] (2, 1.5) rectangle(6, 2.5);
    \node[right] at (6, 2) {\Large
        $f_3^1 : \{A_3^1=1, L_3^1=3\}$
    };

    \fill[Gray] (3, 3) rectangle(6, 4);
    \draw[thick] (2, 3) rectangle(6, 4);
    \node[right] at (6, 3.5) {\Large
        $f_2^1 : \{A_2^1=2, L_2^1=2\}$
    };

    \fill[Gray] (4, 4.5) rectangle(6, 5.5);
    \draw[thick] (2, 4.5) rectangle(6, 5.5);
    \node[right] at (6, 5) {\Large
        $f_1^1 : \{A_1^1=2, L_1^1=2\}$
    };

    \draw[thick] (0.5,0.25) rectangle(1.5,5.25) node[pos=0.5, rotate=270] {\Large Load Balancer};

    \draw[thick, ->] (1.5, 0.5) -- (2, 0.5);
    \draw[thick, ->] (1.5, 2.0) -- (2, 2.0);
    \draw[thick, ->] (1.5, 3.5) -- (2, 3.5);
    \draw[thick, ->] (1.5, 5.0) -- (2, 5.0);

    \draw[thick, ->] (-1.5, 2.75) -- (0.5, 2.75);
    \node[above] at (-0.75, 2.75) {\Large Flows to};
    \node[below] at (-0.75, 2.75) {\Large VIP 1};

\end{tikzpicture}
    \caption{An example of L4 load balancing with 4 DIPs with their load highlighted.
    The load balancer receives two elephant flows in a short span of time.
    }
    \label{example}
\end{figure}

In this section, we first review existing flow dispatchers and
demonstrate their shortcomings.
Then, we introduce a novel L4 flow dispatching algorithm.
Throughout the section, we use a simple example as shown in
Figure~\ref{example} to compare the performance of various flow dispatchers.
In this example, the VIP has four DIPs ($f_1^1$, $f_2^1$, $f_3^1$, $f_4^1$) with available capacities of 2, 1, 0, and 0 units.
We assume that the load balancer receives two elephant flows in a very short span of time.
Flow dispatcher can maximize the throughput by assigning one elephant flow to each $f_2^1$ and $f_1^1$.
We compare the aggregated throughput of flow dispatchers by analyzing their likelihood of assigning
an elephant flow to an already overwhelmed DIP.

\subsection{Existing Flow Dispatchers}

\subsubsection{Equal Cost Multipath Routing}
ECMP is the most commonly used flow dispatcher due to its simplicity.
Under ECMP, a new connection is equally likely to be dispatched
to any DIP in the pool:
\begin{equation*}
    \forall j, \forall i: p[f_i^j] = \frac{1}{N^j}
\end{equation*}

Consider the example of Figure~\ref{example}.
The probability of two new flows being assigned to $f_1^1$ and $f_2^1$ is equal
to $2*\frac{1}{4}*\frac{1}{4}$ or a mere 12.5\%.
In other words, ECMP is 87.5\% likely to assign at least one elephant flow
to an overwhelmed DIP.

As the example shows, statistically distributing an equal number of flows to DIPs
is not likely to result in a balanced distribution of load due to a number of reasons:
\begin{enumerate}[label=(\roman*)]
    \item Connections have huge size discrepancies;
        indeed, it is well-known that flow size distribution in data centers is
        heavy-tailed~\cite{greenberg2009vl2, alizadeh2011data,
        aghdai2013traffic} and it is quite common for ECMP to map several
        elephant flows to the same resource and cause congestion~\cite{al2010hedera}.
    \item DIPs may have different capacities; this is especially true for
        softwarized instances as in virtualized network functions (VNFs).
    \item ECMP does not react to the state of the system (i.e., oblivious load balancing).
        As our example shows, ECMP may dispatch new connections to overwhelmed
        instances and deteriorate $\Omega^j$ as a result.
\end{enumerate}

Recent load balancers~\cite{araujo2018balancing,olteanu2018stateless} use
consistent hashing.
While consistent hashing is an excellent tool for assuring PCC, it aims to achieve
the same goal as ECMP in equalizing the number of assigned flows to DIPs.
These solutions achieve the same performance as ECMP in terms of
load balancing efficacy.
Therefore, we categorize solutions based on consistent hashing in the same
performance class as ECMP.

\subsubsection{LCF}

Least-Congested First (LCF) is a dispatching algorithm mainly used at L3, but
we analyze its performance if applied at L4.
LCF is stateful; it periodically polls instances' utilization and
dispatches new connections to the instance with the least utilization.

For the example of Figure~\ref{example}, LCF considers $f_1^1$
as the least utilized DIP until the next polling; therefore, it dispatches both
of the connections to that instance.
As a result, the two elephant flows are assigned to $f_1^1$, while $f_2^1$ has
available capacity to spare.
In other words, if two elephant flows arrive in a short span of time, LCF is 100\%
likely to assign them into the same DIP.
%Compared to ECMP, LCF relies on polling of resources' utilization to perform
%stateful load balancing.

LCF's performance heavily depends on the polling frequency.
As our example shows, LCF potentially performs
worse than ECMP when too many flows enter the system within a polling cycle. 
LCF-based routing schemes process flowlets~\cite{sinha2004harnessing}
rather than flows and use very short polling intervals ranging from a few
RTTs~\cite{alizadeh2014conga} to O(1ms)~\cite{katta2016hula}.
Therefore, LCF is not suitable for L4 flow dispatching since:
\begin{itemize}
    \item Frequent polling leads to extensive communication and
        processing overhead and hinders scalability.
        %since the number of DIPs is much larger than the
        %number of available paths in L3 - tens of thousands compared to hundreds.
    \item LCF puts a lot of burst load on the least-congested DIP until
        the next polling cycle.
\end{itemize}

\subsection{Adaptive Weighted Flow Dispatching (AWFD)}

AWFD is our proposed stateful flow dispatcher at L4.
It polls DIPs' available capacity to avoid sending new flows to
overwhelmed DIPs.
Within each polling cycle, AWFD distributes new flows among a group of
uncongested DIPs.
Therefore, AWFD reduces the pressure on DIPs and allows for less frequent
polling of instances' status.
Since many DIPs with various available capacities may be active simultaneously,
AWFD assigns weights to DIPs to assure that it dispatches 
incoming flows to instances in proportion to DIPs' available capacity.

We optimize AWFD for implementation on programmable data planes.
The first step is to partition the DIP pool into smaller groups comprised of
DIPs with roughly equal available capacity.
We use \emph{priority classes} (PCs) to refer to such groups of DIPs.
As illustrated in Figure~\ref{awfd-logical}, AWFD breaks down flow dispatching
for new flows into two stages:
\begin{figure}[t]
    \centering
    \pagenumbering{gobble}

\begin{tikzpicture}[every text node part/.style={align=center}]
    
    \draw[thick] (2, 3) rectangle(4, 5) node[pos=0.5] {Choose \\ Priority \\ Class};

    \draw[thick] (6, 0) rectangle(8, 2) node[pos=0.5] {ECMP on \\ Priority \\ Class m};
    \node at(7,2.75) {\Huge .};
    \node at(7,2.50) {\Huge .};
    \node at(7,2.25) {\Huge .};
    \draw[thick] (6, 3) rectangle(8, 5) node[pos=0.5] {ECMP on \\ Priority \\ Class 2};
    \draw[thick] (6, 5.25) rectangle(8, 7.25) node[pos=0.5] {ECMP on \\ Priority \\ Class 1};

    \draw[dashed] (1.65,0) -- (1.65,8);
    \draw[dashed] (4.35,0) -- (4.35,8);
    \node[below] at (3, 8) {\textbf{Stage I}};

    \draw[dashed] (5.65,0) -- (5.65,8);
    \draw[dashed] (8.35,0) -- (8.35,8);
    \node[below] at (7, 8) {\textbf{Stage II}};

    \draw[thick, ->] (0, 4) -- (2, 4);

    \draw[thick, ->] (5, 1.00) -- (6, 1.00);
    \draw[thick, ->] (5, 4.00) -- (6, 4.00);
    \draw[thick, ->] (5, 6.25) -- (6, 6.25);

    \draw[thick] (5, 3.50) -- (5, 1.00);
    \draw[thick] (4, 3.50) -- (5, 3.50);
    \draw[thick] (4, 4.00) -- (5, 4.00);
    \draw[thick] (4, 4.50) -- (5, 4.50);
    \draw[thick] (5, 4.50) -- (5, 6.25);

    \draw[thick, ->] (8, 1.00) -- (9.1, 1.00);
    \draw[thick, ->] (8, 4.00) -- (9.1, 4.00);
    \draw[thick, ->] (8, 6.25) -- (9.1, 6.25);

    \node[above] at (0.75, 4) {Flows to};
    \node[below] at (0.75, 4) {VIP $j$};

    \node[above] at (8.75, 1.00) {To};
    \node[below] at (8.75, 1.00) {DIP};
    \node[above] at (8.75, 4.00) {To};
    \node[below] at (8.75, 4.00) {DIP};
    \node[above] at (8.75, 6.25) {To};
    \node[below] at (8.75, 6.25) {DIP};
\end{tikzpicture}
    \caption{Logical view of AWFD algorithm}
    \label{awfd-logical}
\end{figure}
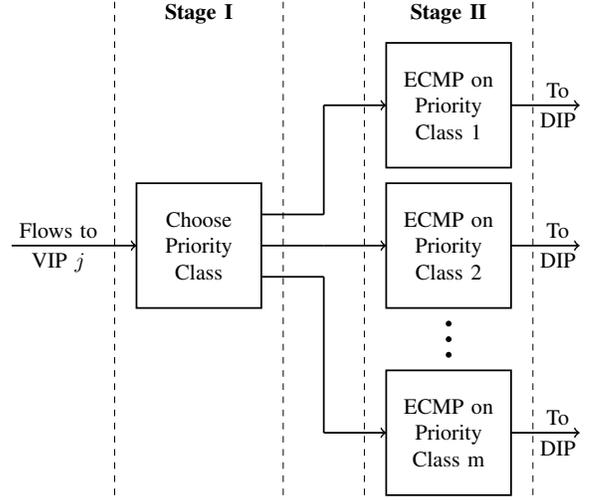

\begin{enumerate}
    \item \textbf{Stage I:} Choose a PC for incoming flows. 
        The probability of choosing a PC is proportional to the
        sum of the available capacities of its members.
    \item \textbf{Stage II:} Assign incoming flows to a DIP from the chosen PC.
        Since members of each PC have an almost equal available
        capacity, AWFD randomly selects one of them to serve the new flow, i.e. same as ECMP.
\end{enumerate}

Next, we formally define AWFD and show that the two-stage selection algorithm
assigns new flows to DIPs in proportion to DIPs' available capacity.

\subsubsection{Design of AWFD}

In this section, we assume that the instances' available capacity are
available to the flow dispatcher.
\S\ref{estimationSubSection} elaborates how Spotlight estimates available
capacities. AWFD is formally defined using the following notation: 
\begin{description}
    \item $m$: Maximum weight assigned to network function instances.
    \item $k$: Weight index for priority classes $(0 \leq k \leq m)$.
    \item $w_i^j$: Weight assigned to $f_i^j$ $(0 \leq w_i^j \leq m)$.
    \item $B_k^j$: PC $k$ for $j$th VIP 
        i.e., set of all instances of $j$th VIP that have weight of $k$.
    \item $||B_k^j||$: Number of instances in $B_k^j$.
    \item $p[B_k^j]$: Probability of choosing PC $k$ for a new connection
        assigned to VIP $j$.
    \item $p[f|B_k^j]$: Probability of choosing DIP $f$ of $B_k^j$ given that
        $B_k^j$ was selected by the first stage of AWFD.
\end{description}

To form PCs, AWFD quantizes DIPs' available capacity into an
integer between $0$ and $m$ and use it as instances' weight:
\begin{equation*}
    \forall j, \forall i: w_i^j =
    \lfloor m \frac{A_i^j}{max_i(A_i^j)} \rfloor
\end{equation*}
Instances with the same weight have an almost equal available capacity and form
a PC.
When a new connection enters the system, the first stage of AWFD assigns it
to a PC with a probability that is proportional to the aggregated
weight of PC members:
\begin{equation}\label{awfdRange}
    \forall j: p[B_k^j] = \frac{
        \sum_{i: f_i^j \in B_k^j}{w_i^j}}{\sum_{i}{w_i^j}} =
    \frac{k||B_k^j||}{\sum_{i}{w_i^j}}
\end{equation}
Note that in the first stage the probability of choosing $B_0$ (the group of DIPs
with little to no available capacity) and empty classes is zero.
Therefore, overwhelmed DIPs of $B_0$ are inactive
and do not receive new connections.
The second stage selects an instance from the chosen non-empty PC with equal
probabilities:
\begin{equation*}
    \forall j, \forall k > 0, \forall f \in B_k^j: p[f | B_k^j] =
    \frac{1}{||B_k^j||}
\end{equation*}
Given that the two stages of the algorithm work independently, we have:
\begin{equation} \label{AWFD-prob}
    \forall j, \forall k, \forall f \in B_k^j: p[f] =
    p[f | B_k^j] p[B_k^j] = \frac{k}{\sum_{i}{w_i^j}}
\end{equation}

Equation~\ref{AWFD-prob} shows that AWFD dispatches new flows to DIPs in
proportion to DIPs' weights.
Since we use DIPs' available capacity to derive their weight, the DIPs
receive new flows according to their available capacity.
Algorithm~\ref{alg:awfd} formally describes AWFD.

\begin{algorithm}[t]
    \begin{algorithmic}
        \Function{AWFD}{5-Tuple flow information}
        \State $ flID \gets Hash(5-Tuple) $ \Comment{5-tuple flow identifier}
        \State $ wSum \gets \sum_i{w_i^j} $
        \If {$ flID \% wSum \leq ||B_1^j|| $}
            \State $B=B_1^j$
        \ElsIf {$ flID \% wSum \leq ||B_1^j||+2*||B_2^j|| $}
            \State $B=B_2^j$
            %\State $.$
            \State $...$ \Comment{Stage I: Choose priority class $B$}
            %\State $.$
        \ElsIf {$ flID \% wSum \leq \sum_{k=1}^{m-1}{k||B_k^j||} $}
            \State $B=B_{m-1}^j$
        \Else
            \State $B=B_m^j$
        \EndIf
        \State \Return {$f \gets B[flID \% ||B||]$} \Comment{Stage II: choose DIP $f$ from $B$}
        %\State \Return $F$
        \EndFunction
    \end{algorithmic}
    \caption{AWFD DIP assignment algorithm for new flows to $VIP^j$}
    \label{alg:awfd}
\end{algorithm}

AWFD scales in data plane as well as in control plane.
In the data plane, all of the operations of the two stages of the algorithm are
compatible with P4 language~\cite{bosshart2014p4} and can be ported to any
programmable switch.
From the control plane point of view, AWFD does not require per-flow rule 
installation.
Instead, forwarding decisions are handled at the switches' data plane when new
connections arrive.
The control plane periodically transfers PC updates to
switches and enables them to make in-band flow dispatching.

AWFD is a general model for flow dispatching.
Existing schemes such as weighted fair queuing (WFQ), ECMP, and LCF are special cases
of AWFD.
If we increase the value of $m$ and the rate of updates, AWFD performance will
be similar to that of WFQ.
Choosing a small value for $m$ likens AWFD to LCF and
AWFD with $m=1$ is equivalent to LCF since all DIPs will be deactivated, apart
from the one with highest available capacity.
AWFD with no updates and $m=0$ is equivalent to ECMP as all DIPs are put into
a single PC regardless of their available capacity and they are equally likely
to receive a new flow.

Consider the example of Figure~\ref{example};
under AWFD with $m=2$, instances get the following weights:
\begin{align*}
    &w_1^1= \lfloor 2*\frac{2}{2} \rfloor = 2\\
    &w_2^1= \lfloor 2*\frac{1}{2} \rfloor = 1\\
    &w_3^1=w_4^1 =\lfloor 2*\frac{0}{2} \rfloor = 0
\end{align*}
Therefore, $f_1^1$, $f_2^1$ will receive 66.6\%, and 33.3\%
of new connections until the next round of polling.
Therefore, the probability of two elephant flows being assigned to 
$f_1^1$, $f_2^1$ is equal to $2*\frac{1}{3}*\frac{2}{3}$ or 44.4\% which is much
better than ECMP and LCF.

By dispatching new flows to multiple instances in different PCs, AWFD reduces
the burstiness of traffic dispatched to instances.
As a result, DIPs' available capacity are less volatile compared to LCF,
and the polling frequency can be reduced as well.
Thus, AWFD is more scalable than LCF as the amount of traffic on the control
plane is reduced.

\subsubsection{Implementation of AWFD in Data Plane}
\begin{figure}[t]
    \centering
    \includegraphics[width=0.495\textwidth]{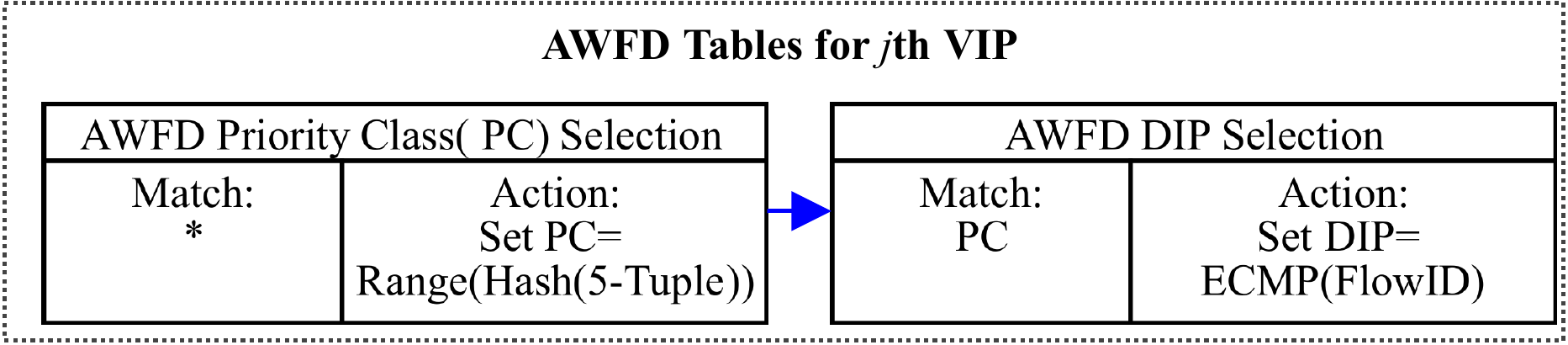}
    \caption{AWFD Tables in data plane.}
    \label{awfdFig}
\end{figure}

AWFD is implemented using P4 language.
As shown in Figure~\ref{awfdFig}, we use two tables for the two stages of
the algorithm.
The first table selects a PC based on the 5-tuple flow identifier.
In Eq.~\ref{awfdRange} we have established the probability of choosing each PC.
Since there are $m$ PCs\footnote{$B_0$ has a probability of 0; therefore we
exclude it from the rest of PCs.} for each VIP, the random selection of PCs
is implemented in the data plane using $m+1$ discrete ranges in
$(0, \sum_{i}{w_i^j})$ as explained in Algorithm~\ref{alg:awfd}.
The first table includes the ranges and utilizes P4 range matching to map
the hash of 5-tuple flow information to one of the ranges and attaches the
matched range as PC metadata to the packet.

The second table, corresponding to the second stage of the algorithm includes
$m$ ECMP groups corresponding to PCs.
This table matches on the PC metadata and chooses one of the ECMP groups 
accordingly.

%From the control plane point of view, $n$ weight updates for the DIPs
%may result in $m$ updates in the first table and $2n$ on the second table as
%the updated DIPs are removed from their ECMP group and added to another.
%After each polling that results in $n$
%DIP weight updates, at most $2n+m$ updates should be performed at each switch.

\begin{figure*}[t]
    \centering
    \begin{minipage}[b]{0.68\textwidth}
        \includegraphics[width=\linewidth]{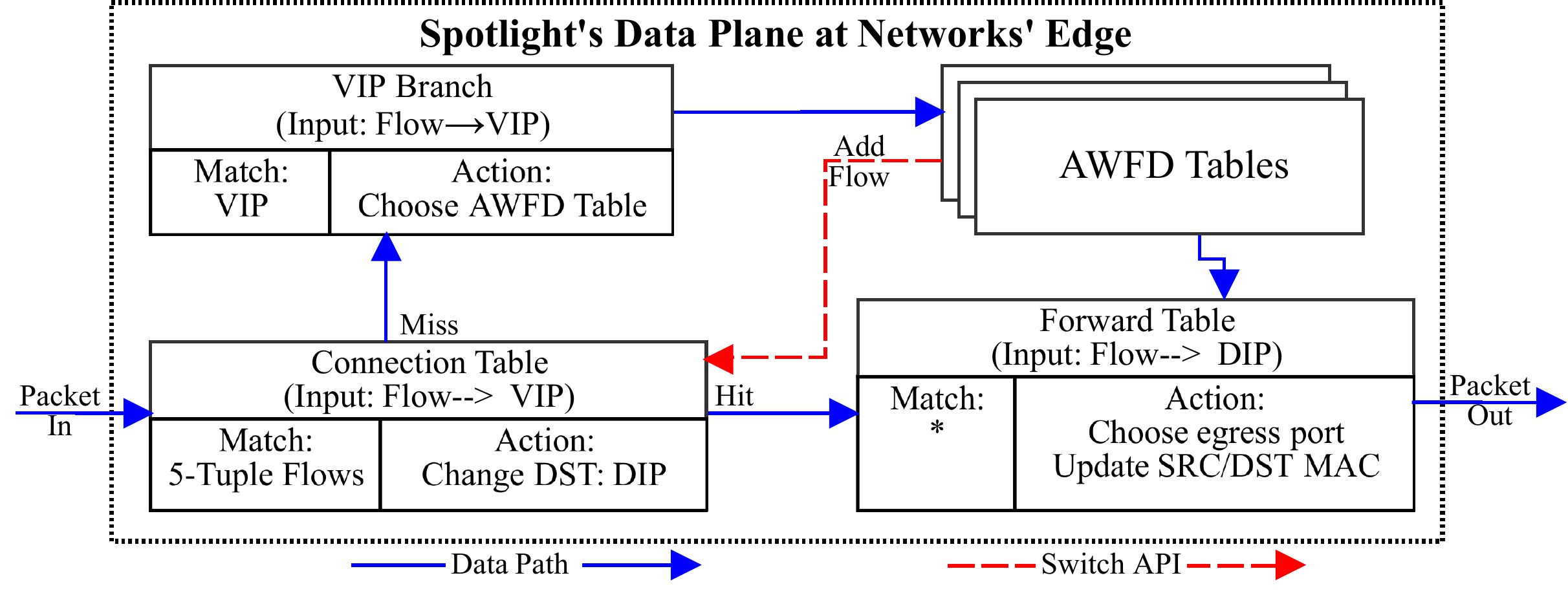}
        \caption{Spotlight's data plane.}
        \label{dataPlaneFig}
    \end{minipage}
    \begin{minipage}[b]{0.31\textwidth}
        \includegraphics[width=\linewidth]{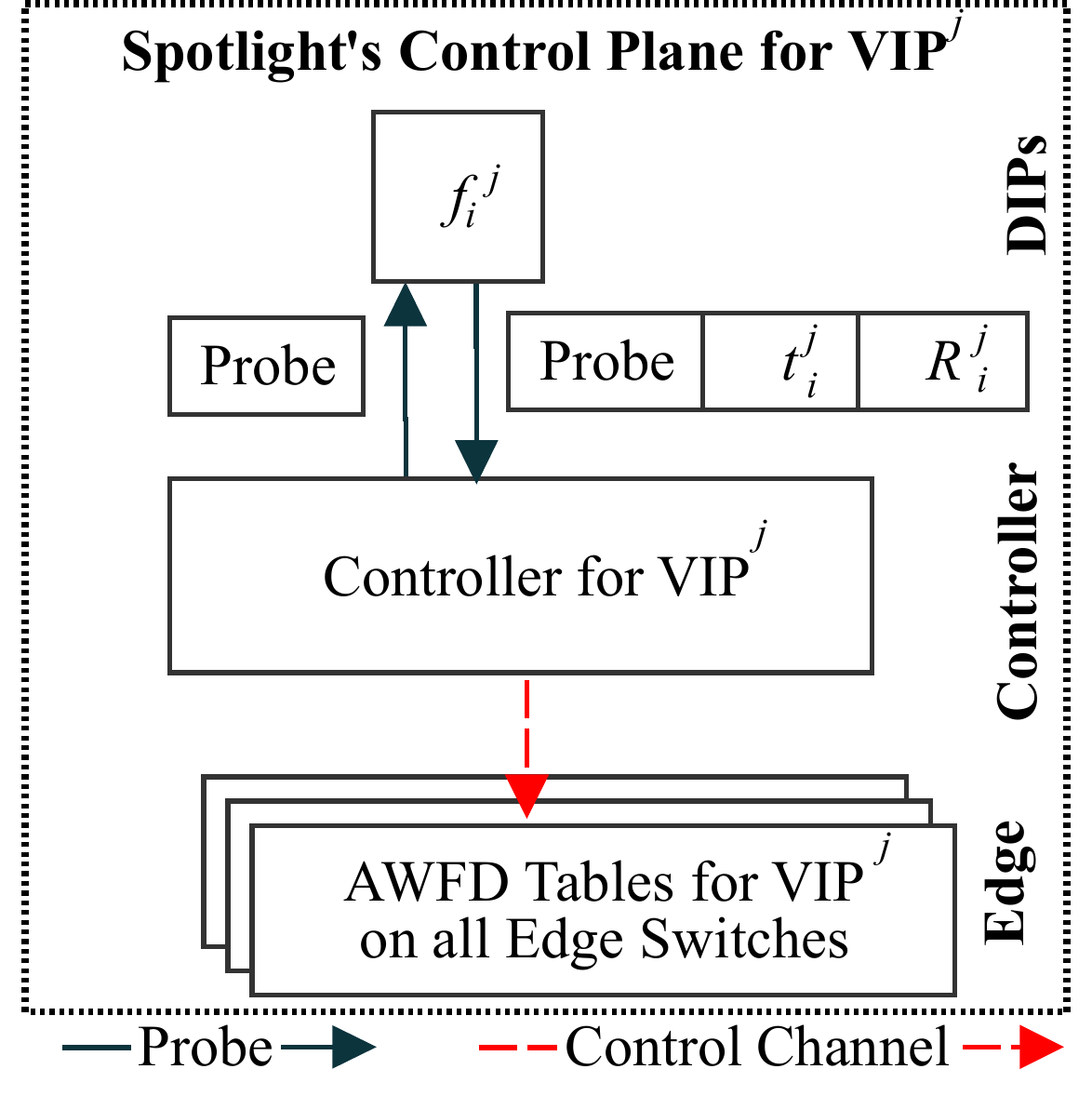}
        \caption{Spotlight's control plane is distributed over VIPs.}
        \label{controlPlaneFig}
    \end{minipage}

\end{figure*}

\section{Spotlight Architecture}\label{SpotlightSec}
Spotlight periodically polls DIPs to estimate their \emph{available capacity}.
During each polling interval, it uses AWFD to distribute new flows among DIPs
in proportion to their available capacity.
Spotlight uses distributed load balancing at connections' source as illustrated
in Figure~\ref{distLB}.
It is implemented at the ~\emph{Programmable edge of the network}, i.e., the
programmable networking device that is located close to connections' source.
Using P4 Language~\cite{bosshart2014p4,p4lang}, we can program the data plane
and port it to a compatible top of the rack (ToR)
switch~\cite{bosshart2013forwarding, tofino,aghdai2017design},
smart network interface card (NIC)~\cite{nfp4000,zilberman2014netfpga,ozdag2012intel,
xpliant}, or software switch at the hypervisor~\cite{shahbaz2016pisces} with little
or no modification.

Spotlight's control plane delivers AWFD weights to edge devices; it is
distributed across VIPs and scales horizontally.
Spotlight flow dispatcher works at flow-level and is decoupled from L3
routing.
Therefore, Spotlight is orthogonal to flowlet-based L3 multi-path routing
schemes and can be implemented on top of such protocols.

\subsection{Data Plane}\label{SpotlightDataPlaneSubSection}
Figure~\ref{dataPlaneFig} illustrates Spotlight's data plane.
The data plane includes a connection table that maps existing connections to
DIPs as well as AWFD tables that contain controller-assigned AWFD ranges and
ECMP groups.
AWFD tables are updated periodically and are used to assign new flows to DIPs
in-band.

VIP-bound packets first pass the connection table.
The connection table guarantees PCC by redirecting existing connections to
their pre-assigned DIPs. 
If the connection table misses a packet, then the packet either
belongs to a new connection, or it belongs to a connection for which the DIP is
assigned at the data plane, but the switch API is yet to enter the rule at the
connection table (discussed in \S\ref{versionSubSub}).
Packets belonging to new flows hit AWFD tables:
the first table chooses a priority class and the second table selects a DIP
member from the chosen class using ECMP.

Similar to Silkroad~\cite{miao2017silkroad}, once a DIP is chosen, a packet
digest is sent to the switch API which adds the new DIP assignment to the
connection table.

\subsection{Estimation of Available Capacity}\label{estimationSubSection}
Spotlight estimates the available capacity of service instances using their
average processing time and load.
As shown in Figure~\ref{controlPlaneFig}, Spotlight polls DIPs to collect
their average processing time ($\overline{t_i^j}$) and load ($L_i^j$).

For each instance, the average processing time is used to estimate 
its capacity:
\begin{equation*}
    C_i^j = 1 / \overline{t_i^j}
\end{equation*}
The available capacity of the DIP is then approximated using its capacity and
load:
\begin{equation*}
    %A_i^j = (1-U_i^j)C_i^j = C_i^j - R_i^j
    A_i^j = C_i^j - L_i^j
\end{equation*}

\S\ref{subSecDis} elaborates how these values can be acquired from the data
plane if DIPs do not report them to the controller.

\subsection{Control Plane}
As shown in Figure~\ref{controlPlaneFig}, the controller is distributed across
VIPs, i.e., each VIP has a dedicated controller.
During each polling cycle, the controller multicasts a probe to DIPs to poll
$\overline{t_i^j}$ and $R_i^j$, and uses these values to approximate instances' 
available capacity.
The controller regularly updates AWFD tables at edge switches.

\subsubsection{Control Plane Scalability}
In addition to distributing the control plane per VIP, Spotlight uses the following
techniques to reduce the amount of control plane traffic and improve scalability.
\begin{itemize}
    \item \textbf{In-band flow dispatching.}
        Spotlight's control plane programs AWFD tables that dispatch new
        connections in the data plane.
        As a result, incoming flows do not generate control plane traffic.
    \item \textbf{Compact AWFD updates.}
        Spotlight controllers only transfer updates in AWFD tables to switches.
        It means that if a controller updates the priority class for $x$ DIPs,
        it has to update at most $m$ ranges in the first AWFD table, remove 
        $x$ DIPs from old ECMP groups, and add them to the new ECMP groups.
        Therefore, for $x$ DIP updates at most $m + 2x$ updates are sent to
        switches.
        Given that $m$ is a small number for AWFD, the number of new
        rules at the switch is proportional to the number weight updates in the
        DIP pool -- a small value at the steady state.
    \item \textbf{Low-frequency AWFD polling.}
        AWFD assigns weights to DIPs according to their available capacity to
        ensure that multiple DIPs are active in each polling interval.
        As a result, AWFD is less sensitive to the polling frequency.
        Spotlight updates switches after every polling.
        Using long polling intervals reduce the amount of control plane traffic.
\end{itemize}

\subsection{Discussion} \label{subSecDis}

\subsubsection{How does Spotlight utilize legacy DIPs that do not
support the reporting of load and processing times to the controller?}

A Programmable data plane can estimate and report average processing time for
legacy DIPs that do not communicate with Spotlight controller.
To measure the average processing time at networks' edge, we can
sample packets from different connections using a Sketch
algorithm~\cite{kumar2006sketch}.
The switch adds an In-band Network Telemetry (INT)~\cite{kim2015band}
header to each sample packet that includes the packets' arrival time and
directs them to the assigned DIP.
After the DIP processes the packet, it returns to the edge switch
that uses current time and the included timestamp in the INT header to estimate
the processing time.
Then, the switch sends the DIP's estimated processing time ($t_i^j$) to the
controller.
The controller uses an Exponential Weighted Moving Average Generator (EWMA) to
estimate the average processing time.
%\begin{equation*}
%    \overline{t_i^j} \leftarrow \alpha t_i^j +
%    (1 - \alpha) \overline{t_i^j}
%\end{equation*}

\subsubsection{What happens if the connection table grows too large to fit in
the limited memory of edge switches?} \label{subsubLargeCon}

There are multiple solutions to this problem:
\begin{enumerate}[label=(\roman*)]
    \item The connection table at the switch may be used as a cache, 
        while an SDN application keeps the complete copy of this table.
        If a packet misses the connection table, it either belongs to a new
        connection, or it belongs to an existing connection not present in the
        cache.
        New connections are identified by checking the TCP SYN flag and are
        processed by AWFD tables.
        For existing connections that miss the connection table cache, a request
        is sent to the SDN application to update the cache.
        
    \item Silkroad~\cite{miao2017silkroad} proposes to limit the size of 
        connection tables by using hash tables on the
        switch to accommodate a complete copy of the connection table.
        The probability of hash collisions can be reduced by utilizing Bloom 
        filters~\cite{bloom70space}.
        
    \item Thanks to the P4 language, Spotlight can easily be ported to a NIC or
        a virtual switch at hypervisor.
        These devices have ample amount of available memory; while a switch may have
        tens to hundreds of MegaBytes of SRAM, NICs and virtual switches have GigaBytes
        of DRAM.
        Moreover, the number of connections at the host or VM level is much smaller
        compared to ToR level.
        Therefore, smart NICs and virtual switches can conviniently host the full copy of
        the connection table.
\end{enumerate}

\subsubsection{How does Spotlight ensure PCC if a load balancer fails?}
If a load balancer fails, its connection table will be lost; therefore, we have
to make sure that other load balancers can recover the connection-to-DIP
mapping to ensure PCC for connections assigned to the failing load balancer.
Any Spotlight instance can recover the connection to DIP mapping within the
same polling interval since both stages of AWFD use 5-tuple flow information
to assign connections to instances.
However, for old connections, the AWFD tables may get updated; as such the
connection-to-DIP mapping cannot be recovered for old connections.

To solve this issue, we have to rely on an SDN application to track
connection tables at Spotlight load balancers.
If a device fails, the connection table at other devices will act as a cache,
and the SDN application can provide connection-to-DIP mapping for the cache
misses (see the first answer to \S~\ref{subsubLargeCon}).

\subsubsection{How does Spotlight guarantee PCC when new connections
are being added to the connection table?}\label{versionSubSub}

The switch API adds new flows to the connection table.
However, in the current generation of P4-compatible devices, the ASIC cannot add new rules.
Therefore, we may observe some delay from the time that the first packet of a flow is processed
in the data plane to the time that the switch API adds the corresponding rule to the connection table.
As such, subsequent packets of the flow may miss the connection table.
AWFD tables use 5-tuple flow information to assign new connections, and therefore 
this delay would not cause PCC violation if AWFD tables are not changed during
the update delay.
However, PCC may be violated if a periodic AWFD update takes place during the
update delay.
Therefore, to meet PCC, we have to guarantee that during the update delay AWFD
tables is not changed.

Silkroad~\cite{miao2017silkroad} proposes a solution to this problem:
Edge switches keep track of multiple versions of flow dispatching tables to
ensure consistency when control plane updates such tables.
For new connections, the data plane keeps track of latest version of the
tables at the time of arrival for the first packet of the flow.
The version metadata is stored in registers updated in the data plane by
the ASIC.
For subsequent packets of the flow, the value of the
register points the data plane to the correct version
of AWFD tables to be used for the flow.

\subsubsection{Control Plane Convergence}\label{sec:disc:control}
Lack of convergence in the control plane is a possible obstacle in scalability
of SDN applications.
Networks are unreliable and provide best effort delivery of messages.
As a result of delayed or lost control messages, switches may end up in an
inconsistent state which may degrade SDN applications' performance or even break
their operation.
This scenario becomes more likely as the SDN application scales out to more
switches.
Spotlight controller broadcasts AWFD weights to switches periodically.
Therefore Spotlight control plane is prone to lack of convergence.

However, the potential inconsistent state among Spotlight load balancers does
not break load balancing as a network function.
Load balancers assign new flows to DIPs based on AWFD weights and add
the new flows to their connection table which is a local state and is 
not synchronized.
In other words, AWFD weights are the sole shared
state among controller and load balancers.
Load balancers can operate with an inconsistent state (AWFD weights) as
they still assign new flows to DIPs and add them to the local
connection tables.

The inconsistent state may potentially degrade the performance
as load balancers that use outdated weights are more likely to assign
incoming traffic to overwhelmed DIPs.
However, due to the closed-loop feedback, Spotlight is highly resilient to
state inconsistencies.
The controller periodically polls instances, updates AWFD weights,
and broadcasts them to switches.
If some switches do not receive the updated weights, they will keep
using the old weights.
Transient weight inconsistencies impact DIPs' state which are monitored by the controller.
The effect of the inconsistency is reflected in the next set of AWFD
weights which will be broadcast to all switches.
Hence, switches with inconsistent state will have the chance to recover.
We observe this process in our testbed and its performance impact is measured
in \S\ref{sec:eval:control}.

\section{Evaluation}\label{evalSec}
% Few points to make
% 1. use table to summarize the simulation setting
% 2. explain what happen when a new flow is assigned to a congested instance
% 3. number of updates needed per sampling period
% 4. address 

\subsection{Flow-level Simulations}
\begin {table}
\caption{Flow Statistics} \label{tab:FlowStats} 
\footnotesize
\centering
  \begin{tabular}{ | c | c | c |  }
      \hline
      Traffic & Pareto & Production \\ 
      Trace & Distribution & Data Center \\ \hline
      Number of Flows & 100,000 & 357,630 \\ \hline
      Avg. Flow Duration & 10s & 33s \\ \hline
      Avg. Flow Inter-arrival Time & 1ms & 2.5ms  \\ \hline
      Avg. Flow Size  & 2 KBps & 50.7 KBps   \\ 
      \hline
  \end{tabular}
\end {table}

Two different traffic traces are used in the simulation.
To evaluate the effectiveness of Spotlight to dispatch flows to
different instances, flow-level simulation is conducted.
As discussed in the previous sections, instantaneous available capacity of each instance
is used as the weight for AWFD. The rationale behind it is to
have all instances reach full utilization roughly around the same time to avoid overwhelming some instances while under utilizing the others.
Therefore, we use the overall service utilization ($\Omega^j$) as the
performance metric to compare AWFD with other schemes.
$\Omega^j$ is defined as the total carried traffic across all instances
divided by the total capacity across all the instances $VIP^j$.
If a flow is assigned to an instance with an available capacity smaller than
the flow rate, flows running on this instance will have reduced rates 
instead of their intended rates. 
This is because all flows assigned to this congested instance would share the capacity.
As a result, overall service utilization will be degraded as part of the
flow traffic demand cannot be served. That being said, an ideal flow dispatcher should be able to minimize reduced flows and provide higher overall service utilization.

\begin{figure}
    \centering
    \includegraphics[width=0.75\linewidth]{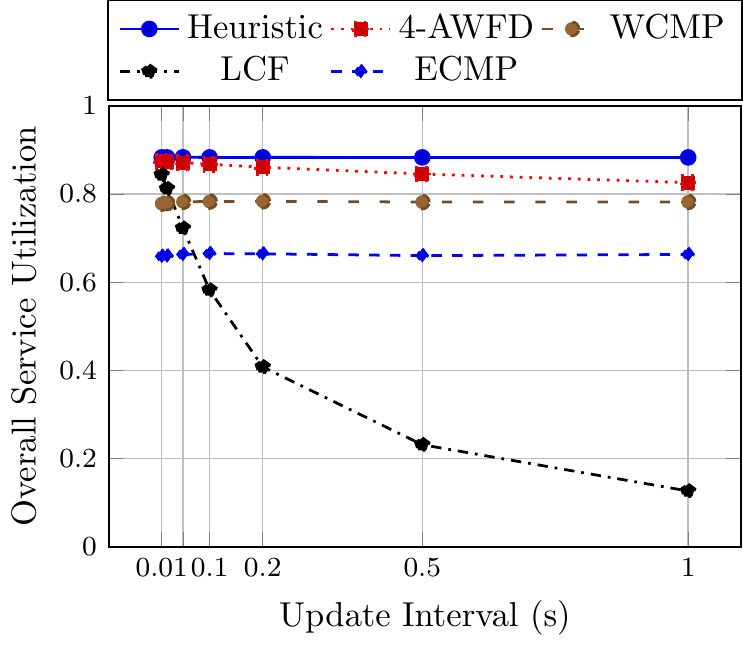}

    \caption{Overall service utilization of different flow dispatchers
    with synthesized trace.}
    \label{fig:synth_result}
\end{figure}

We compare the performance of AWFD to several schemes commonly used in flow dispatchers: ECMP, WCMP and LCF.
ECMP dispatches new flows to all instances with equal probability regardless
of their available capacity.
On the other hand, WCMP assumes the maximum capacity of each instance is known in advance and uses it as a weight to dispatch flows. Therefore, instances with higher maximum capacities have higher chance to receive more flows in proportion to their weights.
As for the LCF, the controller collects the available capacity from all
instances at each update interval and chooses the one with the largest
available capacity.
This instance is then used for all the new flows that arrive before the next
update.
To provide a performance baseline for comparison, heuristic sub-optimal
approach is also used in the simulation.
This approach assumes the controller knows the flow size and the
instantaneous available capacity of each instance when a new flow arrives.
The controller then assigns the new flow to the instance with the
largest available capacity.
This is equivalent to LCF where the updates are done instantaneously. 

The first one is synthesized traffic trace.
We generate the flows based on Pareto distribution to simulate the heavy-tail
distribution found in most data centers~\cite{alizadeh2011data,
greenberg2009vl2,alizadeh2013pfabric,aghdai2013traffic} with the shape
parameter ($\alpha$) set to 2. This heavy-tail distribution provides both mice and 
elephant flows with the numbers of mice flows much more than the elephant flows.
The flow inter-arrival time and flow duration are generated based on Poisson
distribution with the average inter-arrival time set to 1ms and the average
flow duration set to 10 seconds. The flow statistics are summarized in Table \ref{tab:FlowStats}

In total, 100k flows are generated for the simulation.
Four network functions (i.e., VIPs) are used with each having 100 instances
(i.e., DIPs).
Among all the instances, there are two types of DIPs with the a capacity ratio
set to 1:2.
Each flow is assigned to a service chain consisting of 1 to 4 different
services.
The capacity of each instance is configured so that the total requested traffic
is slightly more than the total capacity of the services.
Under this configuration a good flow dispatching scheme can stand out.
Figure \ref{fig:synth_result} shows the overall service utilization of
different schemes, where the x-axis is the update window interval and the
y-axis is the $\Omega$.
As we can see from the figure, AWFD always outperforms both ECMP and WCMP, and its performance
is very close to the heuristic sub-optimal approach.
This is because ECMP does not take into account the available capacity of
the instances.
When there is capacity discrepancy among the instances, ECMP could overflow
those with lower capacity while leaving those with higher capacity
under-utilized.
Although WCMP is able to take into account the maximum capacity discrepancy among the instances and improve the performance over regular ECMP, the lack of knowledge on available capacity makes it still inferior to AWFD. In addition, it is not always feasible to obtain an instance's maximum capacity in advance as it could change dynamically based on other factors such as sharing a resources such as CPU, memory, network interface with other instances that are virtulized on the same physical machine.
The performance of AWFD is very close to LCF when the update interval is very
short.
When the update interval increases, the performance of AWFD only
slightly degrades while the performance of LCF decreases significantly.
This shows that AWFD is less sensitive to the update interval as a
result of using weighted flow dispatching.
On the other hands, LCF is very sensitive to the update interval.
This is because with the longer update interval, the more flows will be assigned to
the least congested instance and could potentially overload it. Besides, it is tricky to choose a proper update interval as it depends on the traffic pattern in the datacenter and the capacity of controllers. 

\begin{figure}
    \centering
    \includegraphics[width=0.75\linewidth]{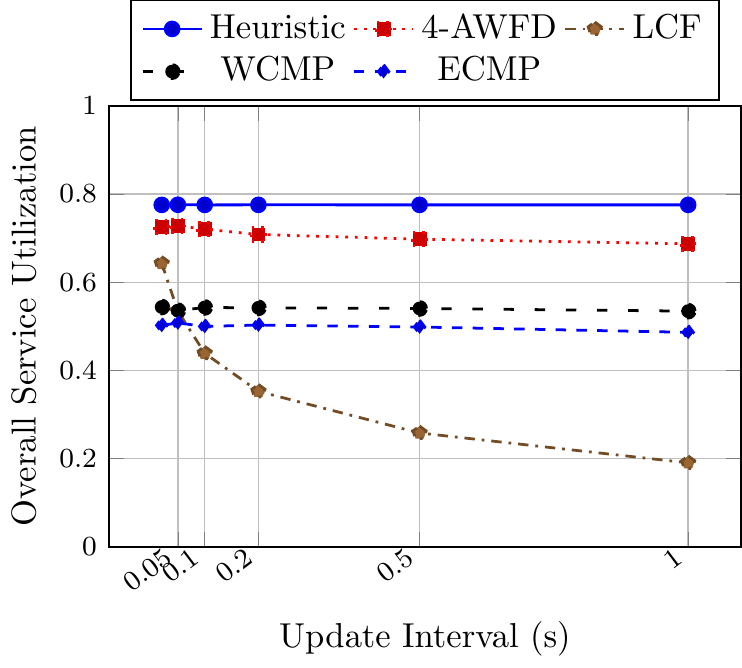}
    \caption{Overall service utilization of different flow dispatchers with
    production data center trace.}
    \label{fig:DCN_result}
\end{figure}

The second traffic trace used for the simulation is backbone traffic from the WIDE MAWI archive~\cite{cho2000traffic}.
There are in total around 357k flows in the captured trace and the duration is
900 seconds.~\footnote{More information on the particular traffic we used is available \href{http://mawi.wide.ad.jp/mawi/ditl/ditl2018/201805091200.html}{here.}}
We have only used flow size, start time and finish
time information, and therefore service chain assignment and instance settings
were synthesized similar to the previous experiment.
Figure \ref{fig:DCN_result} shows the $\Omega$ for different flow dispatchers.
As we can see from the figure, AWFD still outperforms ECMP, WCMP and LCF.
We also observe two differences compared to the previous experiment.
First, the performance of WCMP is much closer to the regular ECMP in this trace.
Second, the performance of AWFD is slightly off the heuristic sub-optimal scheme but much better than WCMP and ECMP.
This is because the flow rate distribution in this trace is not as steep as the synthesized trace which means it has more medium-sized flows.
In order for WCMP to perform well, it requires the majority of flows to be centered around a certain flow size.
Although a more spread-out flow rate distribution has a negative impact on most dispatchers, the impact on AWFD is very minimal.
As discussed in the previous sections, the value of $m$ in AWFD is the
quantization parameter that can be configured and impacts flow dispatching
performance.
With larger $m$ values, the instances are able to obtain weights closer to
their real values based on the available capacity.

\begin{figure}
\centering
\includegraphics[width=0.8\linewidth]{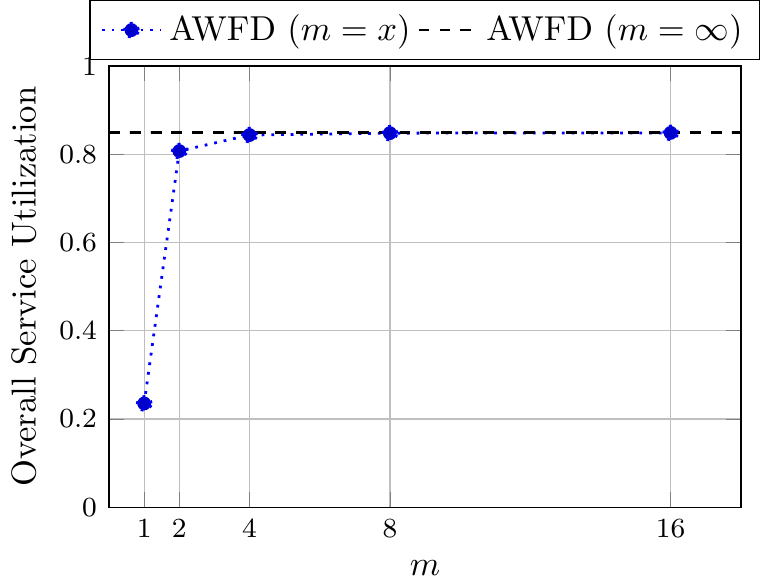}
\caption{AWFD: overall services utilization vs. the value of
    maximum weight ($m$) with the synthesized trace.}
\label{fig:m_value}
\end{figure}

However, larger $m$ values increase the number of priority classes as well as
the number of required updates from the controller and it would increase the
amount of traffic in the control channel.
Therefore, we also evaluate what $m$ value is enough for AWFD to achieve good
performance without burdening the control channel.
With the synthesized trace at update interval of 500ms, we vary
$m$ value from 1 to 16 and compare the performance with $m$ set to infinity,
which means the probability of choosing an instance is exactly proportional to
its available capacity.
Figure \ref{fig:m_value} shows overall service utilization of different
$m$ values.
From this figure, we can see that $m$ values as small as 4 can already
achieve good performance close to the ideal case.

\subsection{Testbed Experiments}
We have implemented Spotlight - including the AWFD flow dispatcher, connection
tables, control plane polling, and periodic AWFD updates - on a small-scale
testbed.
In our experiments, we have used Spotlight to distribute the load among the
instances of a hypothetical file server.
We assume that requests are originated from the Internet and that a distributed
service fulfills the requests.
In our experiments, any instance can serve any request and instances drop the
connections that violate PCC, which is the default behavior of modern OSs.

\begin{figure}
    \centering
    \includegraphics[width=0.2\textwidth]{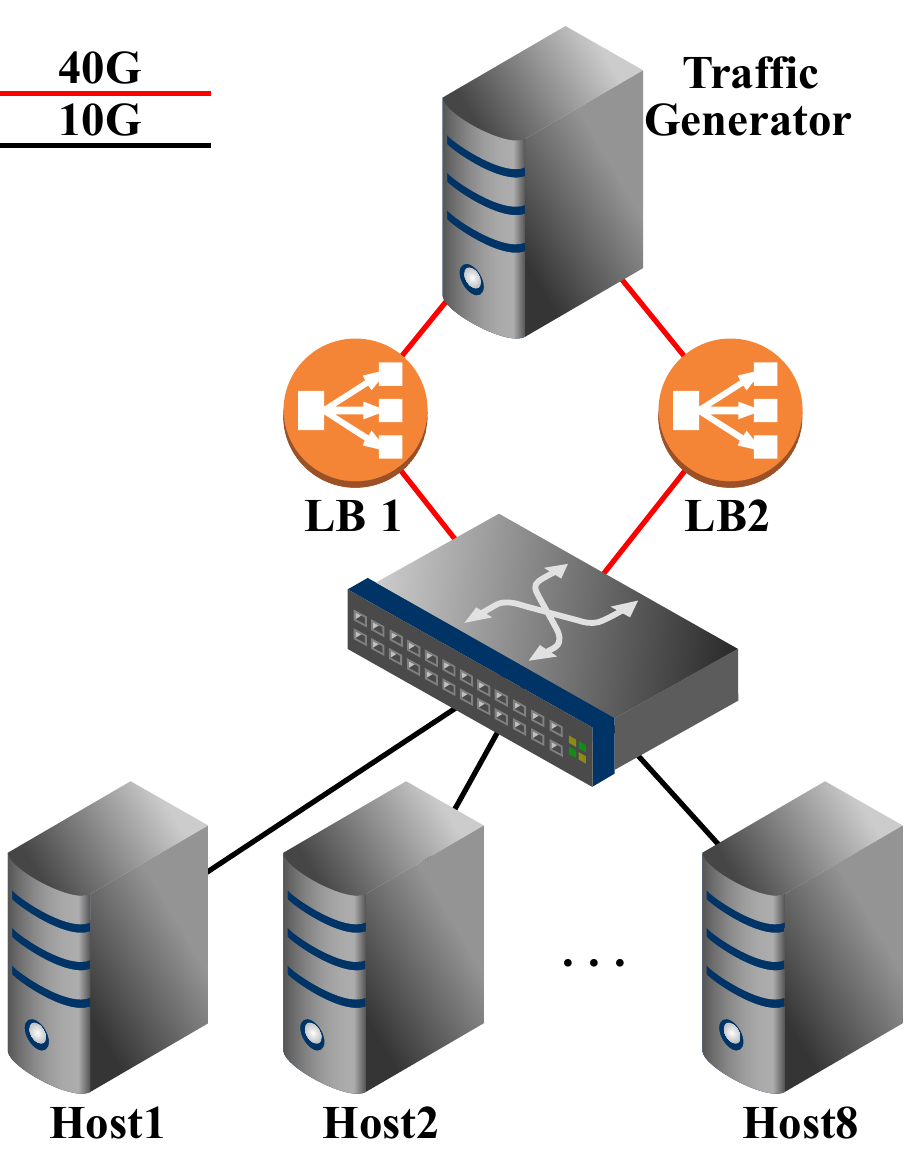}
    \caption{Testbed architecture}
    \label{figTbArch}
\end{figure}

Figure~\ref{figTbArch} illustrates the testbed architecture.
A traffic generator acts as a client that randomly sends requests to
download files.
All of the requests are sent to a VIP, and the client has no
knowledge of DIPs.
Two Spotlight load balancers are configured with the address of the VIP and
are directly connected to two 40G interfaces of the traffic generator.
The traffic generator round robbins the HTTP requests between the two interfaces.
Spotlight load balancers assign incoming connections to DIPs and route them to 40G
uplink interfaces of a switch that connects to all DIPs using 10G Ethernet.
8 servers host 16 DIPs implemented as VMs (2 VM guests per machine).
Servers use quad-core Intel Xeon E3-1225V2 and 16GB of RAM.

The load balancer's data plane is implemented using the
Modular Switch (HyMoS)~\cite{aghdai2017design}.
HyMoS is a platform for testing and rapid implementation of programmable
networks.
It uses hardware, in the form of P4-compatible smart NICs, as well as software
to process packets.
The HyMoS uses dual-port 40G Netronome NFP4000~\cite{nfp4000} NICs on a
server with a 10-core Intel Core i9 7900X CPU and 64GB of RAM.
Spotlight's connection table and AWFD connection tables are implemented on
HyMoS' line cards using P4 language.
AWFD tables and the control plane are implemented on the CPU using a Python
application.
In our implementation, line cards send new flows to the Python application that
runs AWFD.
HyMoS CPU runs the Python application that assigns the new connections to DIPs
and updates line cards' connection tables.

The controller also runs a Python application that polls DIPs' average processing
times to derive AWFD weights.
It also extracts line rate statistics that are used to evaluate the performance.
The controller is implemented on a machine with Intel i9 7900X and 64GB of RAM.

Sixteen DIPs on eight physical machines serve the requests.
As shown in Figure~\ref{figTbArch}, each host runs two DIPs.
DIPs are implemented as common gateway interface (CGI)~\cite{robinson2004common}
applications run on an Apache~\cite{apache} web server.
DIPs drop connections in unknown state~\cite{netstat} -- i.e.,
TCP connections that are not established.
During the course of the experiments that lasted several days, we have not observed
any PCC violations.

In order to avoid storage IO becoming the bottleneck, the CGI application randomly
generates the content of the requested files.
Therefore, in our experiment DIPs' performance is bound by CPU or network IO.
On each host, we have limited the first VMs maximum transfer rate to 3Gbps, and
the second to 2Gbps.
Therefore, the theoretical capacity of the DIP pool is limited at 40Gbps.
However, DIPs' capacities are not constant and fluctuate depending on the number and
size of active connections.
Under heavy loads, with hundreds of open connections, the CPU becomes the bottleneck
and the capacity of the pool drops to less than 40Gbps.

We have run multiple experiments using real traffic traces.
We have used flow size distribution from a production data
center~\cite{aghdai2013traffic} to emulate 50000 files for CGI applications.
For each request, the client randomly chooses one of the files; therefore,
the size of connections in our experiment follows the same distribution
as~\cite{aghdai2013traffic}.
Client requests have a Poisson inter-arrival distribution.
The inter-arrival interval is extracted from~\cite{aghdai2013traffic}.
The event rate ($\lambda$) and the size of flows are multiplied by 20 and 10 respectively
to allow traffic generation at a faster pace.

We have evaluated Spotlight from two aspects: load balancing efficiency and
control plane scalability.

\begin{figure}[t]
    \centering
    \includegraphics[width=\linewidth]{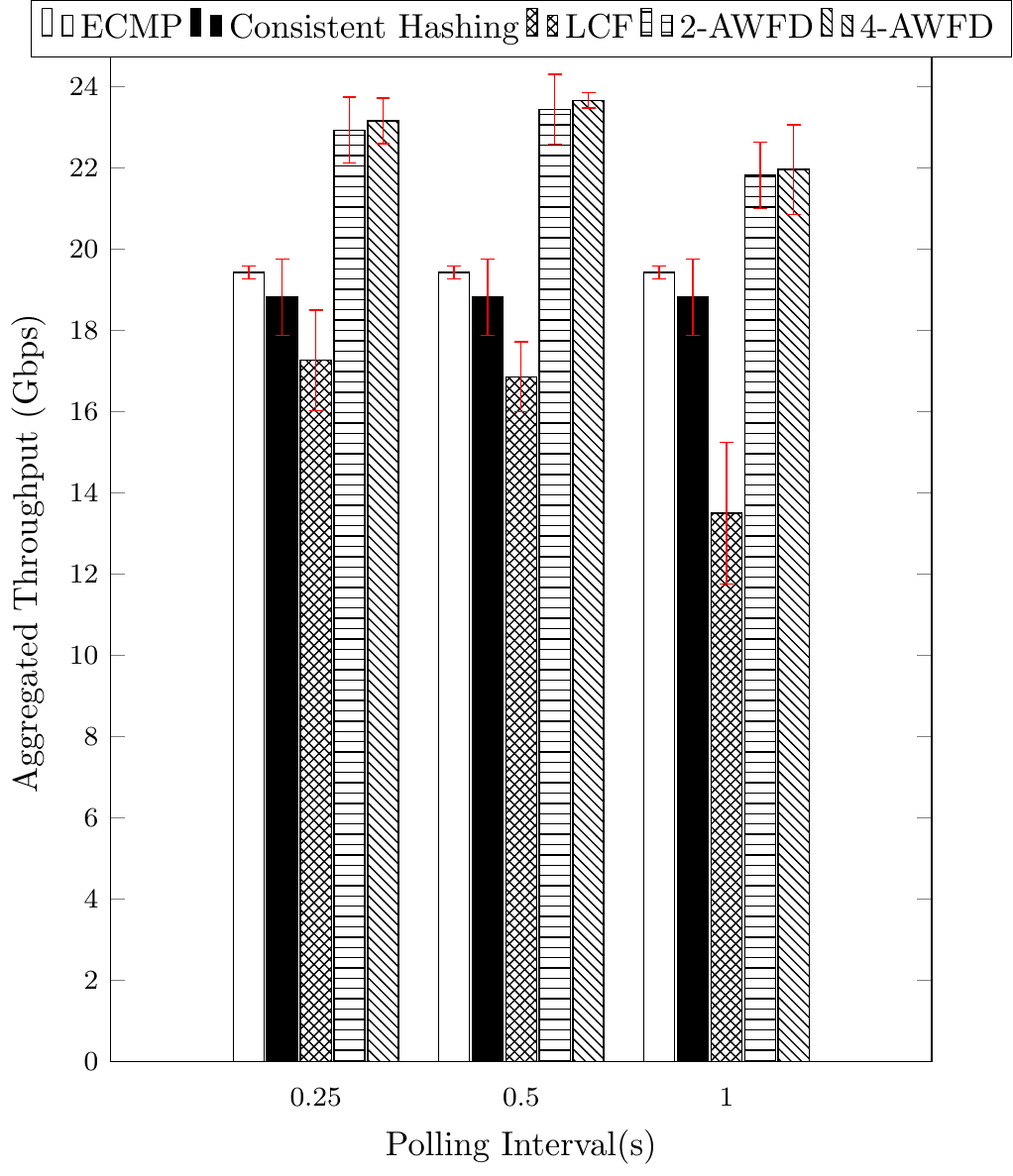}
    \caption{Average overall throughput of different schemes.}
    \label{fig:tput}
\end{figure}

\subsubsection{Performance Measurements}
We compare the performance of Spotlight to state-of-the-art 
solutions~\cite{miao2017silkroad,araujo2018balancing,olteanu2018stateless}
that implement either ECMP or consistent hashing.
Our testbed emulates the two critical conditions of real-world data
centers: heavy-tailed flow size distribution and variable
capacity of DIPs.
Our experiments have two variable parameters: number of AWFD classes ($m$) and
AWFD update interval.

Since the aggregated capacity of our DIP pool is dynamic, we use the aggregated
throughput of DIPs ($T^j$) as the primary performance metric.
We also measured the average completion time of flows which has a high impact
on the responsiveness of the service and users experience.

In the first experiment, we observe the impact of the load balancing algorithm on 
the aggregated throughput of the DIP pool.
In each experiment, the client sends random requests for 60 seconds at a rate
that is close to the maximum capacity of the service.
Experiments are performed 3 times and flow sizes are shuffled in each replication
for more reliable results.
The average value and standard error of measured aggregated throughput are shown in
Figure~\ref{fig:tput}.
Consistent hashing and ECMP flow dispatchers exhibit similar performance.
These methods cannot reach 20Gbps of throughput in our testbed.
However, AWFD with $m=4$ and $500ms$ updates reaches more than 24Gbps of aggregated
throughput on average, a 22\% improvement over ECMP.
AWFD with $m=4$ and $m=2$ consistently outperforms other solutions at $0.25$, $0.5$, and $1s$ update intervals.
AWFD performs slightly worse at $250ms$, due to the high latency of the Python 
controller at the switch that struggles to update the rules in time.
Under $1s$ updates, AWFD performance starts to show higher variance as the
standard error of our measurements increase.
With such large update intervals, having a larger $m$ helps to absorb some of the
negative effects.
LCF ($m=1$) has the worst performance since we are using long update intervals.
Our results show that LCF performance degrades rapidly when the update interval
is prolonged.

\begin{figure}[t]
    \centering
    \includegraphics[width=0.7\linewidth]{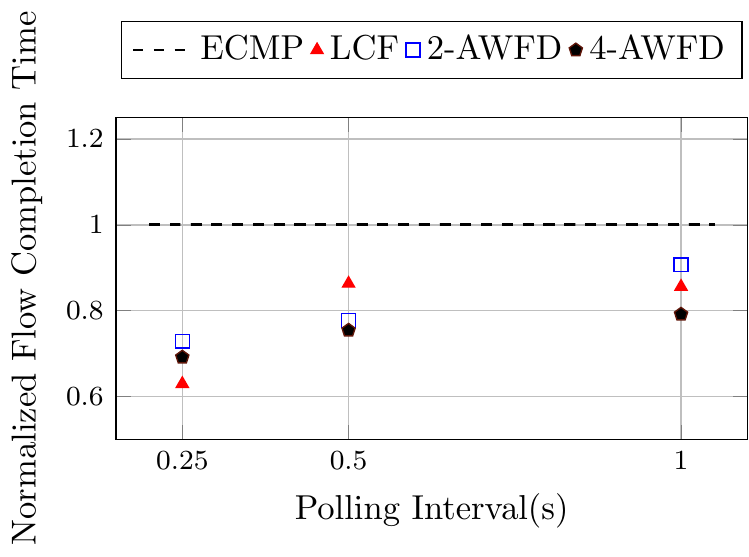}
    \caption{Average flow completion vs. polling interval.}
    \label{fig:fct}
\end{figure}

Next, we use the same settings as the previous experiment to measure
average flow completion time.
As shown in Figure~\ref{fig:fct}, AWFD with $m=4$ shows a consistent improvement
of 20 to 30\% across a range of polling intervals over ECMP.
It is interesting to see that LCF with $250ms$ update interval performs very well under this metric and shows a
37\% improvement over ECMP.
However, the improvement quickly vanishes as we prolong the update interval.
This is because LCF provides superb performance for mice flows
by sending them to DIPs with a high available rate that serve them quickly; however,
it cannot prevent elephant flows from being assigned to the same instance.
Since LCF assigns all flows to the same instance, elephant flows are more likely 
to be routed to the same instance especially at long update intervals.
The assignment of elephant flows heavily impacts the throughput; we observed its impact
in the previous experiment that showed LCF performs poorly in terms of the throughput.
As we expecy, LCF with more frequent polling improves flows average completion time.
However, the improvment margine of LCF becomes smaller as the polling interval increases.
In this context, AWFD is less sensitive to polling frequency compared to LCF; it is
much less likely to send elephant flows to the same DIP due to the usage of weights even at 1s polling interval.
This makes AWFD more capable of delivering high throughput.
On the other hand, using AWFD with short updates, mice flows may still be routed to
overwhelmed instances, and hence, it cannot outperform frequently-polled LCF 
in terms of average completion time.
ECMP and consistent hashing perform poorly in both metrics as they purely
randomize the flow assignment.
In other words, under these schemes elephant flows are more likely to collide
compared to AWFD, and mice flows are less likely to be sent to the least utilized
DIP compared to frequently-polled LCF.

\subsubsection{Control Plane Convergence}\label{sec:eval:control}
Spotlight controller broadcasts AWFD weights to all load balancers making
control plane convergence trivial if messages are delivered to the
switches.
However, networks are unreliable and control messages may get delayed
or lost especially in a scaled-out data center.
In the second set of experiments, we deliberately drop control packets
and observe its impact on the operation of Spotlight.
In other words, scalability is measured in the form of resilience toward 
lack of convergence in the control plane.

As discussed in \S\ref{sec:disc:control}, state inconsistencies do not break
the load balancing function. Neither do they break PCC due to the existence of 
connection tables. However, inconsistencies in AWFD weights may
degrade load balancing performance as they increase the probability of assigning
new flows to congested DIPs.

\begin{figure}[t]
    \centering
    \includegraphics[width=0.75\linewidth]{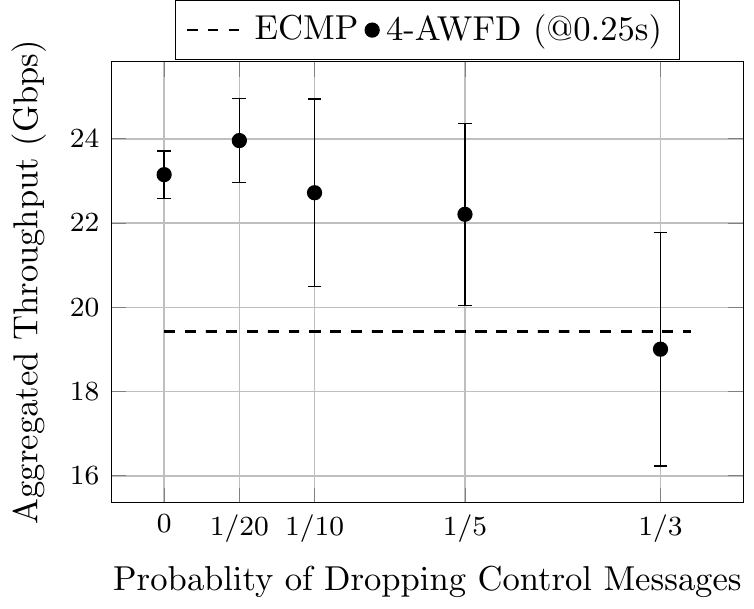}
    \caption{Throughput vs. control message drops.}
    \label{fig:err}
\end{figure}

We measure the impact of inconsistent state using our testbed.
We use the same performance metric in the aggregated throughput of the DIP pool
and define the probability of losing control messages that result in state
inconsistencies as the variable in our experiment.
The experiment is done on the same DIP pool with the same configuration as the
previous section.
File sizes are distributed according to the data center traffic's flow size
distribution, the traffic generator sends requests for 60 seconds with Poisson
inter-arrival times, and experiments are replicated 3 times.

Figure~\ref{fig:err} illustrates the aggregated throughput of AWFD with $m=4$
and $250ms$ polling interval in the vertical axis versus the probability of
dropping control messages by each load balancer on the horizontal axis.
ECMP performance is used as the benchmark.
The results show that under packet drops of less than 20\%, the impact on the
average throughput is negligible.
As we increase the rate of packet drops, the aggregated throughput shows a higher variance with
a much higher standard error.
However, AWFD still outperforms ECMP at 20\% drops.
It is only when we severely increase the probability of drops to an unrealistic
value of one third of messages (33\%) that we observe a significant impact
on AWFD performance.
However, such a scenario is extremely unlikely. Under reasonable assumptions
it is fair to assume that some control messages will be delayed and a small 
percentage of control messages, much less than 10\%, may be dropped.
Such incidents will marginally impact AWFD.
We believe the closed-loop feedback of polling DIPs utilization is the primary
factor in Spotlight's high resilience toward inconvergence of control plane.

\subsubsection{Amount of Control Plane Traffic}
Our results show that Spotlight consistently and reliably outperforms existing
solutions.
From a scalability standpoint, it is interesting to calculate the amount of
control plane traffic using the configuration that we used previously:
$m=4$ and $250ms$ updates.

Assuming that there are $l$ load balancers serving a pool of $n$ DIPs,
controller(s) need to update the weights (a 2 byte number for $m=4$) every
$250ms$ (4 updates per second).
Assuming total packet length of 64 bytes (including packet headers) for control messages,
the total rate of control traffic is equal to $256nl$ Bps.

To put that into perspective, assuming 1000 DIPs and 50 load balancers, the
total amount of control traffic per second would be equal to $12.8MBps$.
It is an insignificant rate of control traffic considering that the data plane 
traffic for this hypothetical DIP pool could easily amount to 1-2Tbps - assuming
a serving capacity of 1-2Gbps per DIP and 20-40Gbps per load balancer.

\section{Related Works}\label{relatedSec}
During the recent years, a number of new load balancers have been proposed for
data center networks.

ECMP-based solutions include Ananta~\cite{patel2013ananta}, an early ECMP-based
software load balancer, Duet~\cite{gandhi2015duet} which introduces hybrid load
balancing by using connection tables in software to maintain consistency while
using ECMP in commodity hardware for flow dispatching, and
Silkroad~\cite{miao2017silkroad} which uses modern programmable data planes for
hardware load balancing with connection tables and ECMP flow dispatching.

Consistent hashing load balancers gained much attention recently.
Maglev~\cite{eisenbud2016maglev} utilizes consistent
hashing~\cite{karger1997consistent} to ensure PCC in face
of frequent DIP pool changes and load balancer failures.
Faild~\cite{araujo2018balancing} and Beamer~\cite{olteanu2018stateless}
implement stateless load balancing using 2-stage consistent hashing;
however, both schemes require some form of cooperation from DIPs to reroute
traffic to the original DIP to maintain PCC when DIP pool is updated.
Therefore, both solutions require modification at hosts' protocol stack to enable rerouting.

Rubik~\cite{gandhi2015rubik} is the only software load balancer that does not
use ECMP;
instead, it takes advantage of traffic locality to minimize the traffic at
data center network's core by sending traffic to the closest DIP.
OpenFlow~\cite{mckeown2008openflow} solutions~\cite{
handigol2009plug,wang2011openflow,kang2015efficient} rely on the
SDN controller to install per-flow, wildcard rules, or a combination of both
for load balancing; while being flexible, per-flow rule installation does not
scale out. Wildcard rules, on the other hand, limit the flexibility of SDN and are
costly to be implemented at TCAM~\cite{yan2018adaptive}.

At L3, however, ECMP-based load balancing has fallen out of favor.
Hedera~\cite{al2010hedera} is one of the earliest works to show ECMP
deficiencies and proposed rerouting of elephant flows as a solution.
CONGA~\cite{alizadeh2014conga} is the first congestion-aware load balancer
that distributes flowlets~\cite{sinha2004harnessing} and prioritizes
least-congested links (LCF).
HULA~\cite{katta2016hula} and Clove~\cite{katta2017clove} extend
LCF-based load balancing on flowlets to heterogeneous networks and at
network's edge, respectively, while having smaller overhead compared to CONGA.
LCF, however, requires excessive polling that ranges from O(Round Trip Time) in
CONGA to O(1ms) in HULA.

WCMP and similar works \cite{zhou2014wcmp,shafiee17simple,rottenstreich2018accurate} 
improve aggregated edge throughput of data centers by assigning flows to
uncongested links by assigning different weights to links.
This family of solutions work at network layer where breaking the flow-route consistency is tolerated as it does not break connections.
On the contrary, Spotlight works at transport layer where violating flow-DIP consistency resets the connections and is unaccapteble.

Spotlight borrows the usage of the connection table from existing work in the field
to meet PCC and combines it with AWFD: a new congestion-aware flow dispatcher
that generalizes LCF and enables less frequent updates.
To the best of our knowledge AWFD is the first in-band weighted congestion-aware flow
dispatcher at L4.

Software Defined Networking~\cite{mckeown2008openflow,gude2008nox,
pfaff2015design,kim2013improving,kang2013optimizing},
Network Function Virtualization~\cite{han2015network,hwang2015netvm,
price2012opnfv},
Programmable switches~\cite{bosshart2013forwarding,jose2015compiling},
and network programming languages such as P4~\cite{bosshart2014p4,p4lang}
are the enablers of research and progress in this area. 
Spotlight, as well as most of the mentioned works in the area, utilize these
technologies to perform load balancing in data centers.

\section{Conclusion}\label{conclusionSec}
In this paper, we take a fresh look at transport layer load balancing in data
centers.
While state-of-the-art solutions in the field define per-connection
consistency (PCC) as the primary objective, we aim to maximize the aggregated
throughput of service instances on top of meeting PCC.

We identify flow dispatching as a performance bottleneck in existing load balancers,
propose AWFD for programmable load-aware flow dispatching and distribute incoming
connections among service instances in proportion to their available capacity.

We introduce Spotlight, as a platform to implement an AWFD-based L4 load
balancer that ensures PCC \emph{and} maximize the aggregated throughput of services.
Spotlight periodically polls instances' load and processing times to estimate
their available capacities and use that to distribute incoming flows among DIPs in proportion to their available capacity.
Distributed control and in-band flow dispatching enable Spotlight's control
plane to scale out while its the data plane scales up with modern programmable
switches.
Through extensive flow-level simulations and testbed experiments, we show
that Spotlight achieves high throughput, improves average flow completion
time, meets the PCC requirement, is resilient to control plane message loss, 
and generates very little control plane traffic.

%In our future work, we aim to combine the benefits of Spotlight
%and consistent hashing and develop a congestion-aware load balancer that meets
%PCC without requiring a connection table.

\bibliographystyle{ieeetr}
\bibliography{references}

\begin{thebibliography}{10}

\bibitem{patel2013ananta}
P.~Patel, D.~Bansal, L.~Yuan, A.~Murthy, A.~Greenberg, D.~A. Maltz, R.~Kern,
  H.~Kumar, M.~Zikos, H.~Wu, {\em et~al.}, ``Ananta: Cloud scale load
  balancing,'' in {\em ACM SIGCOMM Computer Communication Review}, vol.~43,
  pp.~207--218, ACM, 2013.

\bibitem{miao2017silkroad}
R.~Miao, H.~Zeng, C.~Kim, J.~Lee, and M.~Yu, ``Silkroad: Making stateful
  layer-4 load balancing fast and cheap using switching asics,'' in {\em
  Proceedings of the Conference of the ACM Special Interest Group on Data
  Communication}, pp.~15--28, ACM, 2017.

\bibitem{netscaler}
``Netscaler.'' \url{http://www.citrix.com}.

\bibitem{f5}
``F5.'' \url{http://www.f5.com}.

\bibitem{nginx}
``Nginx.'' \url{http://www.nginx.org}.

\bibitem{gandhi2015duet}
R.~Gandhi, H.~H. Liu, Y.~C. Hu, G.~Lu, J.~Padhye, L.~Yuan, and M.~Zhang,
  ``Duet: Cloud scale load balancing with hardware and software,'' {\em ACM
  SIGCOMM Computer Communication Review}, vol.~44, no.~4, pp.~27--38, 2015.

\bibitem{eisenbud2016maglev}
D.~E. Eisenbud, C.~Yi, C.~Contavalli, C.~Smith, R.~Kononov, E.~Mann-Hielscher,
  A.~Cilingiroglu, B.~Cheyney, W.~Shang, and J.~D. Hosein, ``Maglev: A fast and
  reliable software network load balancer.,'' in {\em NSDI}, pp.~523--535,
  2016.

\bibitem{araujo2018balancing}
J.~T. Ara{\'u}jo, L.~Saino, L.~Buytenhek, and R.~Landa, ``Balancing on the
  edge: Transport affinity without network state,'' in {\em 15th USENIX
  Symposium on Networked Systems Design and Implementation (NSDI 18), Renton,
  WA}, 2018.

\bibitem{olteanu2018stateless}
V.~Olteanu, A.~Agache, A.~Voinescu, and C.~Raiciu, ``Stateless datacenter
  load-balancing with beamer,'' in {\em 15th USENIX Symposium on Networked
  Systems Design and Implementation (NSDI)}, vol.~18, pp.~125--139, 2018.

\bibitem{karger1997consistent}
D.~Karger, E.~Lehman, T.~Leighton, R.~Panigrahy, M.~Levine, and D.~Lewin,
  ``Consistent hashing and random trees: Distributed caching protocols for
  relieving hot spots on the world wide web,'' in {\em Proceedings of the
  twenty-ninth annual ACM symposium on Theory of computing}, pp.~654--663, ACM,
  1997.

\bibitem{greenberg2009vl2}
A.~Greenberg, J.~R. Hamilton, N.~Jain, S.~Kandula, C.~Kim, P.~Lahiri, D.~A.
  Maltz, P.~Patel, and S.~Sengupta, ``Vl2: a scalable and flexible data center
  network,'' in {\em ACM SIGCOMM computer communication review}, vol.~39,
  pp.~51--62, ACM, 2009.

\bibitem{al2010hedera}
M.~Al-Fares, S.~Radhakrishnan, B.~Raghavan, N.~Huang, and A.~Vahdat, ``Hedera:
  Dynamic flow scheduling for data center networks.,'' in {\em Nsdi}, vol.~10,
  pp.~19--19, 2010.

\bibitem{alizadeh2014conga}
M.~Alizadeh, T.~Edsall, S.~Dharmapurikar, R.~Vaidyanathan, K.~Chu,
  A.~Fingerhut, F.~Matus, R.~Pan, N.~Yadav, G.~Varghese, {\em et~al.}, ``Conga:
  Distributed congestion-aware load balancing for datacenters,'' in {\em ACM
  SIGCOMM Computer Communication Review}, vol.~44, pp.~503--514, ACM, 2014.

\bibitem{katta2016hula}
N.~Katta, M.~Hira, C.~Kim, A.~Sivaraman, and J.~Rexford, ``Hula: Scalable load
  balancing using programmable data planes,'' in {\em Proceedings of the
  Symposium on SDN Research}, p.~10, ACM, 2016.

\bibitem{katta2017clove}
N.~Katta, A.~Ghag, M.~Hira, I.~Keslassy, A.~Bergman, C.~Kim, and J.~Rexford,
  ``Clove: Congestion-aware load balancing at the virtual edge,'' in {\em
  Proceedings of the 13th International Conference on Emerging Networking
  EXperiments and Technologies}, CoNEXT '17, pp.~323--335, ACM, ACM, 2017.

\bibitem{zhou2014wcmp}
J.~Zhou, M.~Tewari, M.~Zhu, A.~Kabbani, L.~Poutievski, A.~Singh, and A.~Vahdat,
  ``{WCMP:} weighted cost multipathing for improved fairness in data centers,''
  in {\em Ninth Eurosys Conference 2014, EuroSys 2014, Amsterdam, The
  Netherlands, April 13-16, 2014}, pp.~5:1--5:14, 2014.

\bibitem{apache}
``{The Apache Web Server}.'' \url{https://httpd.apache.org/}.

\bibitem{bosshart2013forwarding}
P.~Bosshart, G.~Gibb, H.-S. Kim, G.~Varghese, N.~McKeown, M.~Izzard, F.~Mujica,
  and M.~Horowitz, ``Forwarding metamorphosis: Fast programmable match-action
  processing in hardware for sdn,'' in {\em ACM SIGCOMM Computer Communication
  Review}, vol.~43, pp.~99--110, ACM, 2013.

\bibitem{tofino}
Barefoot, ``Tofino.'' \url{https://www.barefootnetworks.com/technology/}, 2015.

\bibitem{hu2015mobile}
Y.~C. Hu, M.~Patel, D.~Sabella, N.~Sprecher, and V.~Young, ``Mobile edge
  computing—a key technology towards 5g,'' {\em ETSI white paper}, vol.~11,
  no.~11, pp.~1--16, 2015.

\bibitem{att2013domain}
{\relax AT\&T}.~inc., ``A{T}\&{T} domain 2.0 vision white paper,'' 2013.

\bibitem{alizadeh2011data}
M.~Alizadeh, A.~Greenberg, D.~A. Maltz, J.~Padhye, P.~Patel, B.~Prabhakar,
  S.~Sengupta, and M.~Sridharan, ``Data center tcp (dctcp),'' {\em ACM SIGCOMM
  computer communication review}, vol.~41, no.~4, pp.~63--74, 2011.

\bibitem{aghdai2013traffic}
A.~Aghdai, F.~Zhang, N.~Dasanayake, K.~Xi, and H.~J. Chao, ``Traffic
  measurement and analysis in an organic enterprise data center,'' in {\em High
  Performance Switching and Routing (HPSR), 2013 IEEE 14th International
  Conference on}, pp.~49--55, IEEE, 2013.

\bibitem{sinha2004harnessing}
S.~Sinha, S.~Kandula, and D.~Katabi, ``Harnessing tcp’s burstiness with
  flowlet switching,'' in {\em Proc. 3rd ACM Workshop on Hot Topics in Networks
  (Hotnets-III)}, 2004.

\bibitem{bosshart2014p4}
P.~Bosshart, D.~Daly, G.~Gibb, M.~Izzard, N.~McKeown, J.~Rexford,
  C.~Schlesinger, D.~Talayco, A.~Vahdat, G.~Varghese, {\em et~al.}, ``P4:
  Programming protocol-independent packet processors,'' {\em ACM SIGCOMM
  Computer Communication Review}, vol.~44, no.~3, pp.~87--95, 2014.

\bibitem{p4lang}
{\relax The P4 Language Consortium}, ``The p4 language specification.''
  \url{"http://p4.org/spec/"}.

\bibitem{aghdai2017design}
A.~Aghdai, Y.~Xu, and H.~J. Chao, ``Design of a hybrid modular switch,'' in
  {\em Network Function Virtualization and Software Defined Networks (NFV-SDN),
  2017 IEEE Conference on}, p.~6, IEEE, 2017.

\bibitem{nfp4000}
Netronome, ``Nfp-4000 intelligent ethernet controller family.''
  \url{https://www.netronome.com/}.

\bibitem{zilberman2014netfpga}
N.~Zilberman, Y.~Audzevich, G.~A. Covington, and A.~W. Moore, ``Netfpga sume:
  Toward 100 gbps as research commodity,'' {\em IEEE Micro}, vol.~34, no.~5,
  pp.~32--41, 2014.

\bibitem{ozdag2012intel}
{\relax Intel FlexPipe}, ``Intel ethernet switch fm6000 series-software defined
  networking,'' 2012.

\bibitem{xpliant}
{\relax Cavium XPliant}, ``Xpliant packet architecture,'' 2014.

\bibitem{shahbaz2016pisces}
M.~Shahbaz, S.~Choi, B.~Pfaff, C.~Kim, N.~Feamster, N.~McKeown, and J.~Rexford,
  ``Pisces: A programmable, protocol-independent software switch,'' in {\em
  Proceedings of the 2016 conference on ACM SIGCOMM 2016 Conference},
  pp.~525--538, ACM, 2016.

\bibitem{kumar2006sketch}
A.~Kumar and J.~J. Xu, ``Sketch guided sampling-using on-line estimates of flow
  size for adaptive data collection.,'' in {\em Infocom}, 2006.

\bibitem{kim2015band}
C.~Kim, A.~Sivaraman, N.~Katta, A.~Bas, A.~Dixit, and L.~J. Wobker, ``In-band
  network telemetry via programmable dataplanes,'' in {\em ACM SIGCOMM}, 2015.

\bibitem{bloom70space}
B.~H. Bloom, ``Space/time trade-offs in hash coding with allowable errors,''
  {\em Commun. {ACM}}, vol.~13, no.~7, pp.~422--426, 1970.

\bibitem{alizadeh2013pfabric}
M.~Alizadeh, S.~Yang, M.~Sharif, S.~Katti, N.~McKeown, B.~Prabhakar, and
  S.~Shenker, ``pfabric: Minimal near-optimal datacenter transport,'' {\em ACM
  SIGCOMM Computer Communication Review}, vol.~43, no.~4, pp.~435--446, 2013.

\bibitem{robinson2004common}
D.~Robinson and K.~Coar, ``The common gateway interface (cgi) version 1.1,''
  tech. rep., 2004.

\bibitem{netstat}
``Netstat manual page.'' \url{https://www.unix.com/man-page/linux/8/netstat/}.

\bibitem{gandhi2015rubik}
R.~Gandhi, Y.~C. Hu, C.-K. Koh, H.~H. Liu, and M.~Zhang, ``Rubik: Unlocking the
  power of locality and end-point flexibility in cloud scale load balancing.,''
  in {\em USENIX Annual Technical Conference}, pp.~473--485, 2015.

\bibitem{mckeown2008openflow}
N.~McKeown, T.~Anderson, H.~Balakrishnan, G.~Parulkar, L.~Peterson, J.~Rexford,
  S.~Shenker, and J.~Turner, ``Openflow: enabling innovation in campus
  networks,'' {\em ACM SIGCOMM Computer Communication Review}, vol.~38, no.~2,
  pp.~69--74, 2008.

\bibitem{handigol2009plug}
N.~Handigol, S.~Seetharaman, M.~Flajslik, N.~McKeown, and R.~Johari,
  ``Plug-n-serve: Load-balancing web traffic using openflow,'' {\em ACM Sigcomm
  Demo}, vol.~4, no.~5, p.~6, 2009.

\bibitem{wang2011openflow}
R.~Wang, D.~Butnariu, J.~Rexford, {\em et~al.}, ``Openflow-based server load
  balancing gone wild.,'' {\em Hot-ICE}, vol.~11, pp.~12--12, 2011.

\bibitem{kang2015efficient}
N.~Kang, M.~Ghobadi, J.~Reumann, A.~Shraer, and J.~Rexford, ``Efficient traffic
  splitting on commodity switches,'' in {\em Proceedings of the 11th ACM
  Conference on Emerging Networking Experiments and Technologies}, p.~6, ACM,
  2015.

\bibitem{yan2018adaptive}
B.~Yan, Y.~Xu, and H.~J. Chao, ``Adaptive wildcard rule cache management for
  software-defined networks,'' {\em {IEEE/ACM} Trans. Netw.}, vol.~26, no.~2,
  pp.~962--975, 2018.

\bibitem{shafiee17simple}
M.~Shafiee and J.~Ghaderi, ``A simple congestion-aware algorithm for load
  balancing in datacenter networks,'' {\em {IEEE/ACM} Trans. Netw.}, vol.~25,
  no.~6, pp.~3670--3682, 2017.

\bibitem{gude2008nox}
N.~Gude, T.~Koponen, J.~Pettit, B.~Pfaff, M.~Casado, N.~McKeown, and
  S.~Shenker, ``Nox: towards an operating system for networks,'' {\em ACM
  SIGCOMM Computer Communication Review}, vol.~38, no.~3, pp.~105--110, 2008.

\bibitem{pfaff2015design}
B.~Pfaff, J.~Pettit, T.~Koponen, E.~J. Jackson, A.~Zhou, J.~Rajahalme,
  J.~Gross, A.~Wang, J.~Stringer, P.~Shelar, {\em et~al.}, ``The design and
  implementation of open vswitch.,'' in {\em NSDI}, pp.~117--130, 2015.

\bibitem{kim2013improving}
H.~Kim and N.~Feamster, ``Improving network management with software defined
  networking,'' {\em IEEE Communications Magazine}, vol.~51, no.~2,
  pp.~114--119, 2013.

\bibitem{kang2013optimizing}
N.~Kang, Z.~Liu, J.~Rexford, and D.~Walker, ``Optimizing the one big switch
  abstraction in software-defined networks,'' in {\em Proceedings of the ninth
  ACM conference on Emerging networking experiments and technologies},
  pp.~13--24, ACM, 2013.

\bibitem{han2015network}
B.~Han, V.~Gopalakrishnan, L.~Ji, and S.~Lee, ``Network function
  virtualization: Challenges and opportunities for innovations,'' {\em IEEE
  Communications Magazine}, vol.~53, no.~2, pp.~90--97, 2015.

\bibitem{hwang2015netvm}
J.~Hwang, K.~Ramakrishnan, and T.~Wood, ``Netvm: high performance and flexible
  networking using virtualization on commodity platforms,'' {\em IEEE
  Transactions on Network and Service Management}, vol.~12, no.~1, pp.~34--47,
  2015.

\bibitem{price2012opnfv}
C.~Price and S.~Rivera, ``Opnfv: An open platform to accelerate nfv,'' {\em
  White Paper. A Linux Foundation Collaborative Project}, 2012.

\bibitem{jose2015compiling}
L.~Jose, L.~Yan, G.~Varghese, and N.~McKeown, ``Compiling packet programs to
  reconfigurable switches.,'' in {\em NSDI}, pp.~103--115, 2015.

\end{thebibliography}

\end{document}